\documentclass[12pt]{iopart}
\expandafter\let\csname equation*\endcsname\relax
\expandafter\let\csname endequation*\endcsname\relax
\usepackage{amssymb}
\usepackage{amsmath}
\usepackage{amsthm}
\usepackage[numbers,sort&compress,square]{natbib}
\usepackage{bm}
\usepackage{graphicx}
\usepackage{hyperref}
\usepackage[capitalise]{cleveref}

\newcommand{\norm}[1]{\left\lVert#1\right\rVert}

\newtheorem{conjecture}{Conjecture}

\theoremstyle{definition}

\begin{document}

\title{Coherent Structures and Travelling Waves in Spatial Replicators from a Biased Volterra Lattice}

\author{Matthew Visomirski$^1$ and Christopher Griffin$^{2}$}
\address{
	$^1$Department of Physics,
	University of Texas at Austin,
    Austin, TX 78705
    }
\address{
	$^2$Applied Research Laboratory,
	The Pennsylvania State University,
    University Park, PA 16802
    }
    
\eads{\mailto{mv33573@my.utexas.edu}, \mailto{griffinch@psu.edu}}

\begin{abstract} The Volterra lattice is a well-known integrable family that  is also a special class of replicator dynamics and whose members can be put in one-to-one correspondence with the directed cycle graphs. In this paper, we study a variation of the Volterra lattice by introducing a bias term in the replicator interaction matrix. The resulting system can still be put into one-to-one correspondence with the directed cycles, and the dynamics offer one generalisation of the classic rock-paper-scissors evolutionary game. We study the resulting spatial dynamics of this family, showing that travelling wave solutions are present in those dynamics corresponding to the directed 5- and 6-cycles, but not the 4-cycle. Instead, the 4-cycle exhibits a set of stationary solutions that we call `frozen waves' that are similar to but distinct from Turing patterns. This type of solution is also found in the dynamics generated from the directed 6- and 8-cycles. We discuss how these stationary solutions can represent naturally emergent ecological niches in these systems, and offer generalizing conjectures for the existence of both travelling wave solutions and frozen wave solutions in this family of dynamics as a potential program of future investigation.
\end{abstract}

\vspace{2pc}

\maketitle









\section{Introduction} 

Replicator dynamics, first introduced by Taylor and Jonker \cite{TJ78}, have been studied extensively \cite{W97,HS98,HS03,FS16}, especially as high-level models of ecosystems \cite{AL11,GBMA17,MCLM23}. Interestingly, these models intersect those from theoretical physics with the discrete KdV equation \cite{M74,KM75,B88} given by replicator dynamics and with certain tournament dynamics in ecology \cite{PG23,I87} also occurring in the analysis of Schr\"{o}dinger operator \cite{VS93}.

In the simplest setting, a payoff or interaction matrix $\mathbf{A} \in \mathbb{R}^{n \times n}$ is given, and the replicator dynamics are given by the system of equations,
\begin{equation}
\dot{u}_i = u_i\left(\mathbf{e}_i^T\mathbf{A}\mathbf{u} - \mathbf{u}^T\mathbf{A}\mathbf{u}\right),
\label{eqn:GeneralAspatialReplicator}
\end{equation}
with $i \in \{1,\dots,n\}$. Here, $\mathbf{u} = \langle{u_1,\dots,u_n}\rangle$ is a vector and $u_i$ is the proportion of species $i$ present in a population. As such, $\mathbf{u} \in \Delta_{n-1}$ where
\begin{equation*}
    \Delta_{n-1} = \left\{\mathbf{u} \in \mathbb{R}^n : \sum_i u_i = 1, u_i \geq 0\right\},
\end{equation*}
is the $n-1$ dimensional unit simplex embedded in $\mathbb{R}^n$. Hofbauer \cite{HS98, HS03}, Weibull \cite{W97} and Tanimoto \cite{T15,T19} provide a complete introduction to evolutionary games and the replicator dynamics.

In the case when $\mathbf{A}$ is skew-symmetric, we have $\mathbf{u}^T\mathbf{A}\mathbf{u} = 0$ and the resulting replicator dynamics simplify to,
\begin{equation}
\dot{u}_i = u_i\mathbf{e}_i^T\mathbf{A}\mathbf{u}.
\label{eqn:AspatialReplicator}
\end{equation}
When $\mathbf{A} \in \mathbb{R}^{n \times n}$ is skew-symmetric and has entries drawn exclusively from $\{0,\pm 1\}$, the resulting matrix is (uniquely) generated by a directed graph $G_\mathbf{A} = (V,E)$, where $V = \{1,\dots,n\}$ and
\begin{equation}
    (i, j) \in E \iff A_{ij} = -1,
    \label{eqn:GraphToMatrix}
\end{equation}
indicating that species $i$ loses to (or is consumed by) species $j$.

Directed cycle graphs give rise to the integrable Volterra lattice (discrete KdV equation) already mentioned \cite{M74,KM75}, but more recent work by several authors \cite{I87,I08,DEKV17,EKV21,EKV22,PG23,VG24} has shown large families of graphs that produce Liouville-Arnold integrable Hamiltonian dynamics along with graph operations that preserve integrability. Simultaneously, rock-paper-scissors and its five-strategy variant, ``rock-paper-scissors-Spock-lizard'' \cite{Kass12} are examples of complete balanced tournaments, whose evolutionary dynamics have also been studied extensively \cite{ML75,RMF07,RMF08,M10,HMT10,SMR13, SMR14,SMJS14,PR17,PR19,KT21,GSB22,PR22}. These games have skew-symmetric payoff matrices (when no payoff bias of the type in \cite{GMD21} is assumed) and in which the corresponding directed graphs are complete with each vertex having equal in- and out-degree (i.e., they are balanced).  It is worth noting that the work in \cite{PR19,M10,SMR14,SMJS14,RMF07,ML75,RMF08,HMT10,PR17,SMR13,KT21,P22,D23} does not use the replicator dynamic and has an alternative competition model for studying cyclic games of this kind.

Spatial variations of the replicator were first studied by Vickers \cite{V89} who added a diffusion term to the standard replicator giving,
\begin{equation}
    \frac{\partial u_i}{\partial t} = u_i\left(\mathbf{e}_i^T\mathbf{A}\mathbf{u} - \mathbf{u}\mathbf{A}\mathbf{u}\right) + D_i\nabla^2 u_i.
    \label{eqn:nDVickers}
\end{equation}
Since then, evolutionary dynamics using partial differential equations have been studied by several authors, with \cite{CV97,dB13,KRFB02,NM92,RCS09,SSI04,V89,V91, HMT10,SMR14,PR17,PR19,GMD21} providing a small example of that body of work.   It is worth noting that Vickers formulation assumes an infinite population, or at least a population in which the total population size is constant in space and time. Durrett and Levin \cite{DL94} were the first to study replicator models where this is violated, and a general non-homogeneous population spatial replicator is derived by Griffin, Mummah and DeForest in  \cite{GMD21} and used to study travelling waves in rock-paper-scissors, which will turn out to correspond to a biased directed 3-cycle in our formulation. The relation between the replicator dynamics of the Prisoner's dilemma game, the Fisher-KPP equation, non-homogeneous populations, and travelling waves is discussed by Griffin in \cite{G23}. Travelling waves in higher-order interactions in both homogeneous and non-homogeneous population spatial replicators are studied by Griffin, Feng and Wu \cite{GFW24}. Contrasting with the work in \cite{PR17,PR19,PR22,P22,D23} in which travelling waves in tournament game dynamics are produced by heteroclinic cycles, the existence of travelling waves in variations of the spatial replicator seems to arise from the presence of a Hopf bifurcation (producing a limit cycle) in the system of ordinary differential equations that are constructed to identify travelling wave solutions \cite{GMD21,GFW24}, except for the Prisoner's Dilemma case, which is fully described by its relation to the Fisher-KPP equation \cite{G23}. Hence, travelling waves in that case are subsumed by the extensive work on the Fisher-KPP equation (see, e.g., \cite{G23,F37,KPP37,PZ04,CSP18,CD17}). 

In this paper, we study both travelling and frozen wave solutions occurring in the spatial replicator equation arising from cyclic graphs; i.e., those spatial replicator equations arising from a variant of Volterra lattice. Following work in \cite{GMD21,GFW24}, we introduce a bias term to the skew-symmetric symmetric matrix defining the original systems in question. Sufficient conditions for the existence of travelling waves are then formulated in terms of the diffusion constant and this bias term. We then show that spatially frozen waves may arise in systems with an even number of species. Using derived results on the 4-, 5-, 6- and 8- cycles we derive general conjectures about this family of spatial dynamics. 

The remainder of this paper is organized as follows: In \cref{sec:aspatial} we introduce the biased Volterra lattice, derived from the directed cycle graphs, and derive stability properties of their fixed points, which we use in our analysis of the spatial case. We discuss the existence of travelling wave solutions in the spatial replicator equations derived from the biased Volterra lattice in \cref{sec:TravellingWaves}. In \cref{sec:FrozenWaves}, we present results on stationary ``frozen wave'' solutions corresponding to the evolution of ecological niches arising in the spatial dynamics of even directed cycles.  and generalize them to other even cycles. Conclusions and future directions are presented in \cref{sec:Conclusion}.

\section{Aspatial Analysis: Biased Volterra Lattice}\label{sec:aspatial}
\begin{figure}[htbp]
\centering
\includegraphics[width=0.7\textwidth]{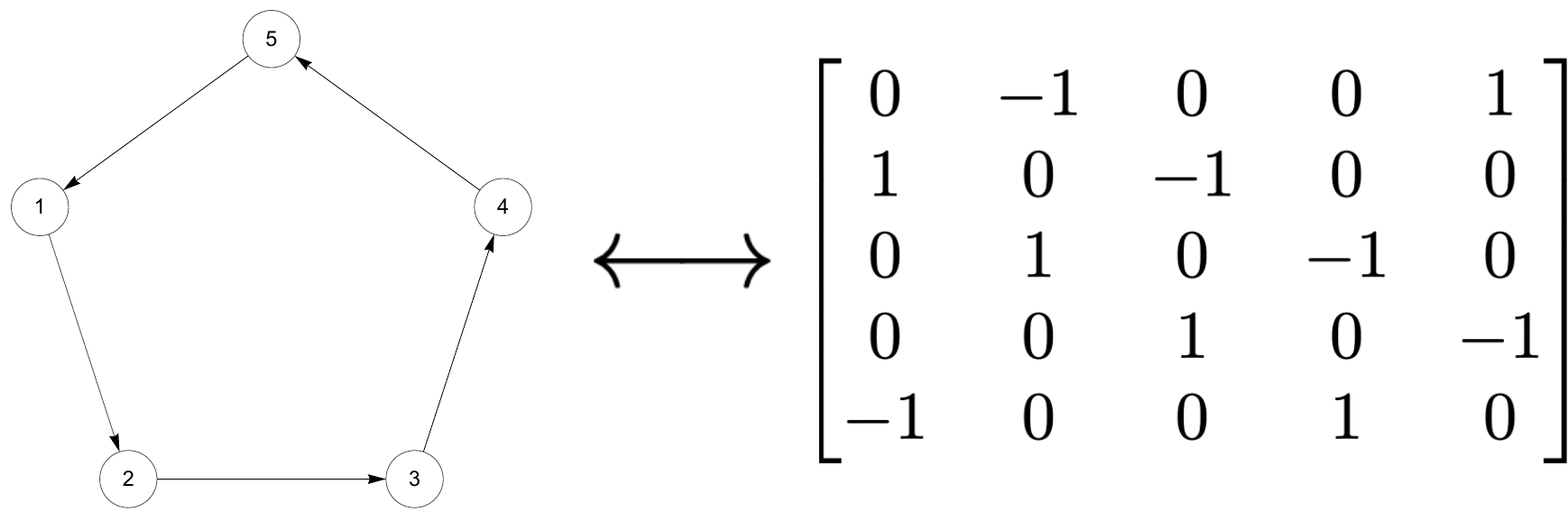}
\caption{The visual relationship between the graph structure and interaction matrix shows how a cycle maps to a matrix, which in turn produces the dynamics.}
\label{fig:FiveCycle}
\end{figure}

The Volterra lattice is the family of differential equations of form,
\begin{equation*}
    \dot{u}_i = u_i(u_{i-1} - u_{i+1}),
\end{equation*}
where addition and subtraction on indices is taken modulo $n$ (for fixed $n$), which imposes periodic boundary conditions. These equations are replicator equations with skew-symmetric interaction matrix,
\begin{equation*}
    \mathbf{A} = \begin{bmatrix}
    0 & -1 & 0 & \cdots & 0 & 1\\
    1 & 0 & -1 & \cdots & 0 & 0\\
    \vdots & \vdots & \vdots & \ddots & \vdots & \vdots\\
    -1 & 0 & 0 & \cdots & 1 & 0
    \end{bmatrix},
\end{equation*}
which, in turn, can be derived from the directed cycle with $n$ vertices using \cref{eqn:GraphToMatrix}, as illustrated in \cref{fig:FiveCycle}. 
As noted, each Volterra lattice (for fixed $n$) is Liouville-Arnold integrable, and the dynamics admit a single elliptic interior fixed point $\mathbf{u}^* = \left\langle\tfrac{1}{n},\dots,\tfrac{1}{n}\right\rangle$ and unstable hyperbolic fixed points on the boundary of the (unit) simplex on which the dynamics evolve. Visomirski and Griffin \cite{VG24} characterized the behaviour of the integrability of these systems by defining non-cycles. A non-cycle is a sequence $(u_{i_1},u_{i_2},\cdots,u_{i_k})$ so that the vertices form a cycle in the graph complement of the directed cycle. A non-edge is defined similarly. For example, $(u_1,u_4)$ is a non-edge of the (directed) five-cycle. While these structures were used to construct conserved quantities in \cite{VG24}, in the sequel we will use these combinatorial objects to construct fixed points. 

If we replace each instance of $+1$ with $1 + a$ in the interaction matrix $\mathbf{A}$, we obtain a biased Volterra lattice defined by the system of differential equations,
\begin{equation*}
    \dot{u}_i = u_i(u_{i-1} - u_{i+1}) - a u_i\left[C_n(\mathbf{u}) - u_{i-1}\right],
\end{equation*}
where,
\begin{equation*}
    C_n(\mathbf{u}) = \sum_{i=0}^{n-1} u_iu_{i+1}.
\end{equation*}
Again, index addition and subtraction is considered modulo $n$. We assume $-1 < a < \infty$ to maintain the cyclic relation given in the graph. 

Direct computation shows that this system also admits the interior fixed point $\mathbf{u}^* = \left\langle\tfrac{1}{n},\dots,\tfrac{1}{n}\right\rangle$ as well as additional hyperbolic boundary fixed points. The stability of any fixed point may be dependent on $a$. It is worth noting the abuse of notation: in theoretical analysis variables are indexed from $0$ to $n-1$ for simplicity in modular arithmetic, but in practice species are often indexed from $1$ to $n$. 

In addition to the interior fixed point, combinatorial reasoning shows that each non-edge generates a fixed point as follows: if $(u_{i_1},u_{i_2},\cdots,u_{i_k})$ is a non-cycle of the underlying graph, then choose $(a_{i_1},\dots,a_{i_k}) \in \Delta_{k-1}$ (i.e., $k$ values between $0$ and $1$ that add to $1$). If $u_j$ is not an element of the non-cycle (or non-edge), then $u_j = 0$. Otherwise, $u_{i_j} = a_{i_j}$. These fixed points are potentially stable depending on the sign of $a$, as we discuss below. Finally, each extreme point $\mathbf{u}^* = \mathbf{e}_i$ (for $i \in \{1,\dots,n\}$), is a fixed point. Here $\mathbf{e}_i$ is the $i^\text{th}$ standard unit normal vector. For the cycles we consider in our spatial analysis, we characterize their fixed points and stability in the following brief discussion in order to explain how Hopf bifurcation and thus travelling waves emerge in the corresponding spatial dynamics.

\subsection*{Four Cycle}
Suppose $\mathbf{F}(\mathbf{u};a)$ is the nonlinear right-hand-side of the biased Volterra lattice generated by the four cycle. At the fixed point $\mathbf{u}^*$ given by $u_1^* = u_2^* = u_3^* = u_4^* = \tfrac{1}{4}$, the eigenvalues of the corresponding Jacobian matrix $\nabla\mathbf{F}$ are given by,
\begin{align*}
    &\lambda_{1,2} = -\frac{a}{4}\\
    &\lambda_{3,4} = \frac{i}{4}(2+a).
\end{align*}
For $a > 0$, this suggests a non-trivial stable manifold (when $a > 0$) with corresponding non-trivial centre manifold, which follows from Theorem 13.3 of \cite{Verh06} and the fact that,
\begin{equation*}
    \lim_{\mathbf{u} \to \mathbf{u}^*} \frac{\norm{\mathbf{F}}}{\norm{\mathbf{u} - \mathbf{u}^*}} = 0,
\end{equation*}
since $\mathbf{F}$ is a nonlinear polynomial expression. Thus, the dynamics near this fixed point are Lyapunov stable, but not asymptotically so, with orbits tending to a centre manifold and then oscillating. This is illustrated in \cref{fig:FourCycle} (left). When $a < 0$, this fixed point is unstable (has non-trivial unstable manifold). 

Alternate fixed points all have form either: $u_1 = u_3 = 0$ and $u_2 = r$ and $u_4 = 1-r$ for some $0 \leq r \leq 1$ or $u_2 = u_4 = 0$ and $u_1 = r$ and $u_3 = 1-r$. These correspond to the two non-edges of the four cycle. In the case when $r = 0$ or $r = 1$ (the extreme point case), the eigenvalues of the corresponding Jacobian matrix are,
\begin{equation}
\lambda_{1,2,3,4} = \{0,0,-1,1+a\},
\end{equation}
implying an unstable hyperbolic fixed point (nonlinear saddle) based on our assumptions on $a$ (i.e., $a > -1$).

\begin{figure}[htbp]
\centering
\includegraphics[width=0.45\textwidth]{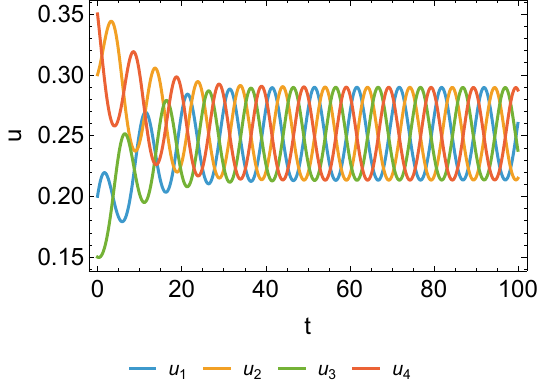} \quad
\includegraphics[width=0.45\textwidth]{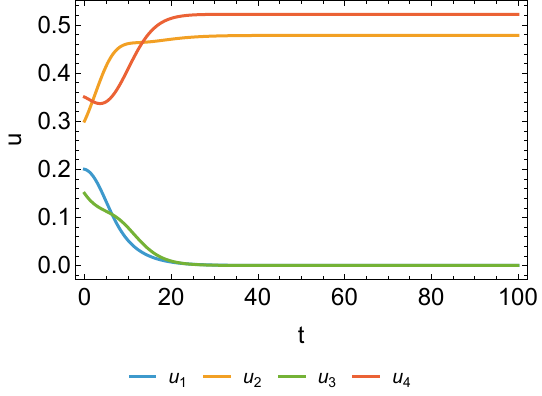}
\caption{(Left) Example dynamics of the biased Volterra lattice on four species with $a = \tfrac{1}{2}$. (Right) Example dynamics of the biased Volterra lattice on four species with $a = -\tfrac{1}{2}$.}
\label{fig:FourCycleExample}
\end{figure}

When $0 < r < 1$, the eigenvalues of the corresponding Jacobian matrix are
\begin{align*}
    &\lambda_{1,2} = 0\\
    &\lambda_{3} = 1 + a - (2+a) r\\
    &\lambda_{4} = -1 + (2+a)r.
\end{align*}
The two zero-eigenvalues correspond to a slow manifold and occur because the dynamics are now restricted to a face of the unit simplex $\Delta_3$; i.e., species cannot spontaneously generate in the absence of mutation. The fixed point is asymptotically stable if (and only if) $a < 0$ and,
\begin{align*}
    \frac{1+a}{2 + a}  < r < \frac{1}{2+a}.
\end{align*}
This is illustrated in \cref{fig:FourCycleExample} (right). These two classes of stable fixed points have a substantial effect on the dynamics of the solutions of the corresponding spatial equations.

\subsection*{Five Cycle}
The analysis for the fixed points of the five cycle is similar, with some variations. As before, let $\mathbf{F}(\mathbf{u},a)$ be the right-hand-side of the biased Volterra lattice generated by the five cycle. Consider the interior fixed point, $u_i^* = \tfrac{1}{5}$ for $i \in \{1,\dots,5\}$. The eigenvalues of $\nabla\mathbf{F}(\mathbf{u}^*)$ are,
\begin{align*}
    &\lambda_1 = -\frac{a}{5}\\
    &\lambda_{2} = -\frac{1}{20}(1+\sqrt{5})a\pm i\frac{2+a}{10}\sqrt{\frac{1}{2}(5 - \sqrt{5})}\\
    &\lambda_{4} = \frac{1}{20}(-1+\sqrt{5})a \pm i\frac{2+a}{10}\sqrt{\frac{1}{2}(5 + \sqrt{5})}.
\end{align*}
Thus the fixed point is a nonlinear saddle and thus unstable.

\begin{figure}[htbp]
\centering
\includegraphics[width=0.45\textwidth]{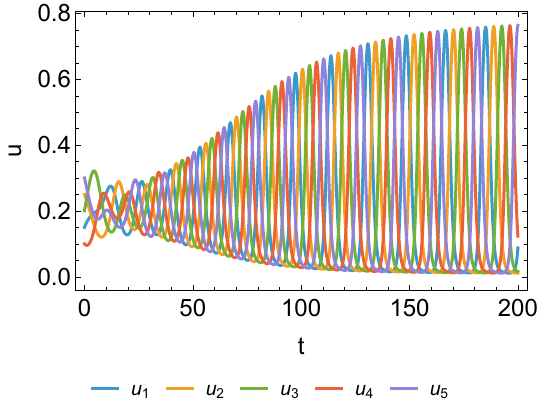}\quad
\includegraphics[width=0.45\textwidth]{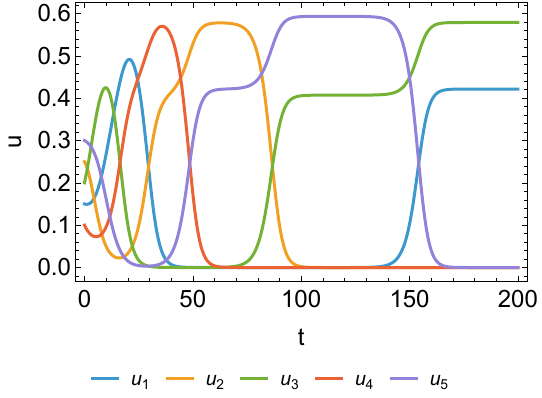}
\caption{(Left) Example dynamics of the biased Volterra lattice on five species with $a = \tfrac{1}{2}$. (Right) Example dynamics of the biased Volterra lattice on five species with $a = -\tfrac{1}{2}$. Note, plots are clipped so that $0 \leq t \leq 200$ at which point the dynamics do not change in character.}
\label{fig:FiveCycleExample}
\end{figure}

Alternate fixed points have the form $u_i = r$ and $u_j = 1-r$ with $0 \leq r \leq 1$ for some pair $(i,j)$ that is not an edge in the five cycle. When $r = 1$ (or $r = 0$) we have a unit vector fixed point. The eigenvalues have form,
\begin{align*}
    &\lambda_1 = -1\\
    &\lambda_2 = 1 - r\\
    &\lambda_{3,4,5} = 0,
\end{align*}
suggesting that the fixed point is unstable and any flow leaving the fixed point if fixed on the boundary, which gives rise to the slow manifolds corresponding to the zero eigenvalues. When $0 < r < 1$ a more interesting phenomenon emerges. At a fixed point of this type, the eigenvalues of $\nabla\mathbf{F}$ are,
\begin{align*}
    &\lambda_{1,2} = 0\\
    &\lambda_3 = -r\\
    &\lambda_4 = (1+a)(1-r)\\
    &\lambda_5 = (2+a)r - 1.
\end{align*}
This fixed point seems unstable and hyperbolic because by our assumption on $a$, $(1+a)(1-r) > 0$. However, for $a < 0$, the directions of instability and stability (given by the corresponding eigenvectors) become important in determining stability and numerical examples show that these fixed points can exhibit a kind of asymptotic stability off the slow and unstable manifolds, just as in the biased four cycle case. This is shown in \cref{fig:FiveCycleExample}.
This behaviour is surprising, but suggests that trajectories remain in the interior of the unit simplex only if $a \geq 0$ with $\omega$-limit sets given by sets of boundary points corresponding to non-edges in the cycle. Even more surprising is the behaviour when $a > 0$. Although the interior fixed point is not a nonlinear centre, the trajectories oscillate as though it is. This can be explained by the instability of the boundary fixed points, the saddle-point nature of the interior fixed point and the fact that straightforward computation, 
\begin{equation*}
    \nabla\cdot\mathbf{F} = a[1 - 7(u_1u_2+u_2u_3 + u_3u_4 + u_4u_5+u_1u_5)] < -\frac{3a}{4} < 0,
\end{equation*}
if $a > 0$. This balance between dissipation in the dynamics and instability in the interior and boundary fixed points combined with the symmetry of the system leads to the observed oscillation. Understanding the exact structure of these $\omega$-limit sets is left for future research, but these phenomena will lead directly to spatial dynamics in the sequel.

\subsection*{Six Cycle}
Analysis of the fixed points of the biased 6-cycle is similar to the analysis of the biased four and five cycles. As before, let $\mathbf{F}$ denote the nonlinear vector field on the right-hand-side of the biased Volterra lattice. At the interior fixed point $\mathbf{u}^* = \left\langle{\tfrac{1}{6},\dots,\tfrac{1}{6}}\right\rangle$, the eigenvalues of the Jacobian $\nabla\mathbf{F}$ are,
\begin{align*}
    &\lambda_{1,2} = -\frac{a}{6}\\
    &\lambda_{3,4} = \pm\frac{a}{12} + i\frac{a+2}{4\sqrt{3}}\\
    &\lambda_{5,6} = \pm\frac{a}{12} - i\frac{a+2}{4\sqrt{3}}.
\end{align*}
It is immediately clear that for $a \neq 0$, this fixed point is always a nonlinear saddle and hence unstable.

At the fixed point $\mathbf{u}^* = \mathbf{e}_i$, the eigenvalues of the Jacobian $\nabla\mathbf{F}$ are,
\begin{align*}
    &\lambda_{1,2,3,4} = 0\\
    &\lambda_5 = 1 + a\\
    &\lambda_6 = -a
\end{align*}
with the zero eigenvalues corresponding to the slow manifold of the dynamics near the extreme point (because each species consumes and is consumed by one other species). 

At the fixed points corresponding to the non-edges of the cycle, for which we have $r \in (0,1)$ so that $u_i = r$, $u_j = 1 - r$ and $u_k = 0$ for $k \neq i,j$ and $(i,j)$ forming a non-edge, the eigenvalues of the Jacobian $\nabla\mathbf{F}$ are,
\begin{align*}
    &\lambda_{1,2,3} = 0\\
    &\lambda_4 = (1+a)(1-r)\\
    &\lambda_5 = -r\\
    &\lambda_6 = (2+a)r - 1.
\end{align*}
This is similar to the five cycle case, in which we had a nonlinear saddle that (surprisingly) was (effectively) stable for certain values of $r$ when $a < 0$. 

\begin{figure}[htbp]
\centering
\includegraphics[width=0.31\textwidth]{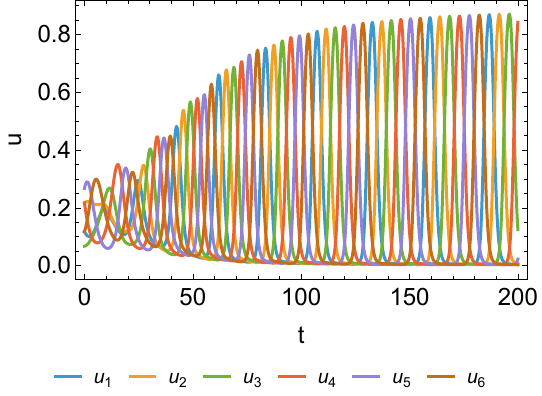}\quad
\includegraphics[width=0.31\textwidth]{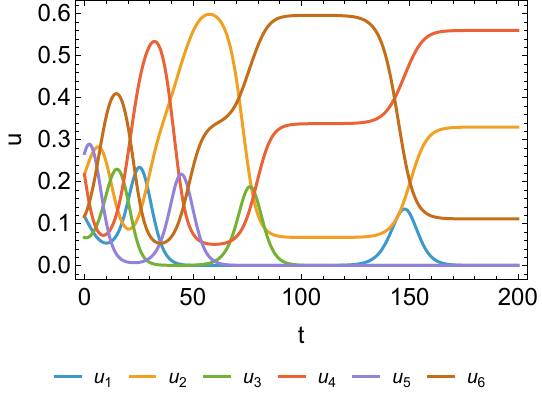}\quad
\includegraphics[width=0.31\textwidth]{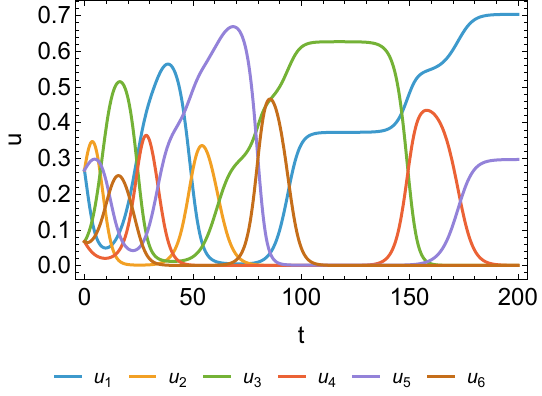}
\caption{(Left) Example dynamics of the biased Volterra lattice on six species with $a = \tfrac{1}{2}$. (Centre) Example dynamics of the biased Volterra lattice on six species with $a = -\tfrac{1}{2}$, settling at a fixed point corresponding to a non-cycle. (Right) Example dynamics of the biased Volterra lattice on six species with $a = -\tfrac{1}{2}$, settling at a fixed point corresponding to a non-edge. Note, plots are clipped so that $0 \leq t \leq 200$ at which point the dynamics do not change in character.}
\label{fig:SixCycleExample}
\end{figure}

For a fixed point corresponding to a non-cycle, that is $p$ and $q$ with $u_i = p$, $u_j = q$, $u_k = 1- p - q$, $u_l =0$ for $l \neq i,j,k$ and $(i,j,k)$ forming a non-cycle of the six-cycle, we have eigenvalues,
\begin{align*}
    &\lambda_{1,2,3} = 0\\
    &\lambda_{4} = p + a p - q,\\
    &\lambda_5 = 1 - 2 p - q + a (1 - p - q),\\
    &\lambda_6 =  (2 + a) q + p -1.
\end{align*}
For $-1 < a < 0$, a sufficient condition for this fixed point to have no unstable manifold is $-1 < a < 0$ and
\begin{equation*}
   \frac{a^2+2 a+1}{a^2+3 a+3}<p\leq \frac{a+1}{a^2+3 a+3} \quad \text{and} \quad \frac{a (-p)+a-2 p+1}{a+1}<q<\frac{1-p}{a+2},
\end{equation*}
which is readily satisfied (e.g.) by $p = q = \tfrac{1}{3}$. As in the case of the four-cycle, we expect trajectories to be drawn to combinations of species corresponding to non-edges or non-cycles of the directed cycle. This is illustrated in \cref{fig:SixCycleExample}.
Again we see dynamics similar to those in both the five cycle and four cycle cases and again, when $a > 0$, we see periodic motion that can be explained by a combination of dissipation and instability. When we analyse the spatial case, we will see that the dynamics arising from these various fixed points will be (partially) stabilized by the diffusion term, leading to both travelling waves and stationary frozen waves that are not explained as Turing patterns.

\section{Derivation of Travelling Waves Solutions in Cycles with Bias}\label{sec:TravellingWaves}

Let $\mathbf{A}$ be the biased-cycle interaction matrix (for an arbitrary cycle) assuming that diffusion is constant over species. In one-dimension, Vicker's equation becomes,
\begin{equation}
    \frac{\partial u_i}{\partial t} = u_i\left(\mathbf{e}_i^T\mathbf{A}\mathbf{u} - \mathbf{u}\mathbf{A}\mathbf{u}\right) + D\frac{\partial^2 u_i}{\partial x^2}.
    \label{eqn:OneDVickers}
\end{equation}
We now show that travelling waves emerge in one-dimension in the spatial variation of the 5- and 6-cycles with bias, leaving the 4-cycle with bias for later. The case when $n = 3$ has been analysed extensively, with travelling waves in this system studied in \cite{GMD21}. Before proceeding, it is worth noting that in finite population models, when $a < 0$, the net population is shrinking. For the case when $n = 3$, Griffin et al. showed that $a < 0$ (i.e., shrinking populations) was a prerequisite for travelling waves to emerge in \cref{eqn:OneDVickers}. For this reason, Griffin, Feng and Wu \cite{GFW24} added higher-order interactions to stabilize the population size while maintaining travelling waves in the 3-cycle (rock-paper-scissors). We will show in the sequel that the requirement that $a < 0$ to see travelling wave solutions is unique to the biased 3-cycle.

To show the existence (or non-existence) of a travelling wave, let $z = x\pm ct$ so that $u_i(x,t) = u_i(z) = u_i(x\pm ct)$. \cref{eqn:OneDVickers} simplifies to the system of ordinary differential equations,
\begin{equation}
    \left\{ \begin{aligned} 
  Dv_i' = cv_i - u_i (e_i^T\textbf{A}u - u^T\textbf{A}u)\\
  u_i' = v_i
\end{aligned} \right.
\label{eqn:TravellingODE}
\end{equation}
Here $c$ represents the wave speed of the travelling wave. Notice this system inherits the fixed points of the aspatial system with the addition that $v_i = 0$ for all $i$. If \cref{eqn:TravellingODE} has a periodic solution, then our original system has a travelling wave solution. We now show that such periodic solutions exist in the various cases by finding an explicit wave speed that induces a Hopf bifurcation and a limit cycle. 





\subsection*{Results for the 5 Cycle}
For the biased 5-cycle, computing the eigenvalues of the Jacobian of \cref{eqn:TravellingODE} evaluated at the interior fixed point $u_i = \tfrac{1}{5}$, $v_i = 0$ for $i \in \{1,\dots,5\}$ yields the eigenvalues,
\begin{align*}
    &\lambda_{1,2}  = \frac{5 c\pm\sqrt{5} \sqrt{4 a D+5 c^2}}{10 D}\\
    &\lambda_{3,4} = \frac{5 c\pm\sqrt{5} \sqrt{5
   c^2-\sqrt{5} a D+a D+i \sqrt{2 \left(5+\sqrt{5}\right)} (a+2) D}}{10 D}\\
   &\lambda_{5,6} = \frac{5 c\pm\sqrt{5} \sqrt{5
   c^2-\sqrt{5} a D+a D-i \sqrt{2 \left(5+\sqrt{5}\right)} (a+2) D}}{10 D}
\end{align*}
\begin{align*}
   &\lambda_{7,8} = \frac{5 c\pm\sqrt{5} \sqrt{5 c^2+\sqrt{5} a D+a D+i \sqrt{10-2 \sqrt{5}} (a+2) D}}{10 D}\\
   &\lambda_{9,10} = \frac{5 c\pm\sqrt{5} \sqrt{5 c^2+\sqrt{5} a D+a D-i \sqrt{10-2 \sqrt{5}} (a+2) D}}{10 D},
\end{align*}
when we assume that $-1 < a < \infty$ and $D > 0$. Note we may write $\lambda_{8}$ and $\lambda_{10}$ as,
\begin{equation*}
    \lambda_{8,10} = \frac{5c - \sqrt{5} \sqrt{5 c^2+\sqrt{5} a D+a D\pm i \sqrt{10-2 \sqrt{5}} (a+2) D}}{10 D}.
\end{equation*}
Assume there is some real $b$ so that,
\begin{equation*}
    \left(\sqrt{5}c \pm bi\right)^2 = 5 c^2+\sqrt{5} a D+a D \pm i \sqrt{10-2 \sqrt{5}} (a+2) D.
\end{equation*}
For this to hold, we must have,
\begin{equation*}
    c^\star = \pm \frac{(a+2) \sqrt{a \left(D-\frac{3 D}{\sqrt{5}}\right)}}{2 a} \quad b^\star = \mp\sqrt{-\left(\sqrt{5} + 1\right) a D},
\end{equation*}
which is valid if and only if $a < 0$. Treating the eigenvalues as a function of wave speed, we see,
\begin{equation*}
    \lambda_{8,10}(c^\star) = \pm\frac{i \sqrt{\frac{1}{5}+\frac{1}{\sqrt{5}}} a}{2 \sqrt{-a D}}.
\end{equation*}
Differentiating with respect to $c$, direct computation shows that, $\lambda_{8,10}'(c^\star) \neq 0$. Thus treating $c$ as a bifurcation parameter, by Hopf's theorem, this system must exhibit a Hopf bifurcation with an emergent limit cycle. This periodic orbit implies the existence of travelling waves for $a < 0$, just as in the 3-cycle case \cite{GMD21}. 

\begin{figure}[htbp]
\centering
\includegraphics[width=0.98\textwidth]{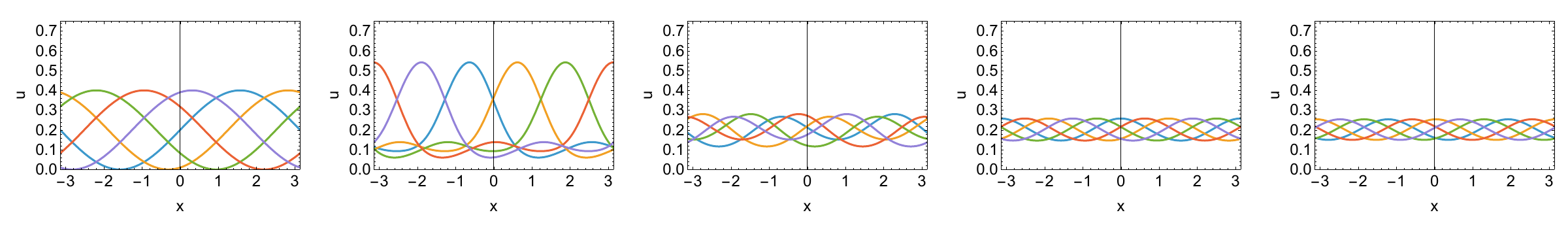}\\
\includegraphics[width=0.98\textwidth]{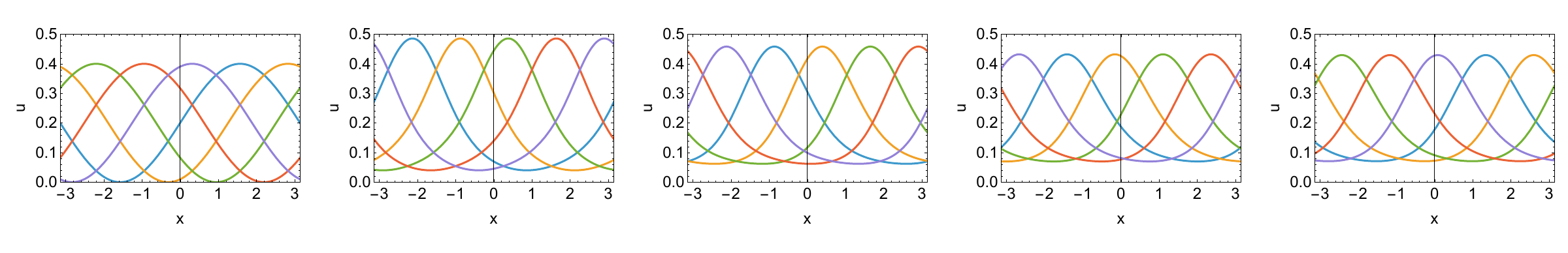}\\
\includegraphics[width=0.98\textwidth]{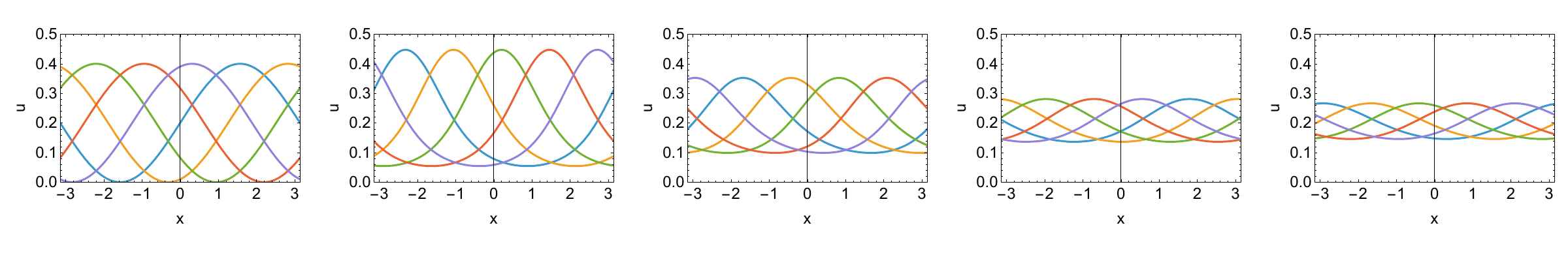}
\includegraphics[width=0.20\textwidth]{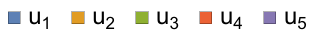}
\caption{(Top) Emergence of a travelling wave with $a = -\tfrac{1}{2}$ and $D = \tfrac{1}{50}$. (Middle) Emergence of a travelling wave with $a = \tfrac{1}{2}$ and $D = \tfrac{1}{50}$. (Bottom) Emergence of a travelling wave with $a = \tfrac{1}{2}$ and $D = \tfrac{1}{30}$. Notice the amplitude of the travelling wave is a function of both $a$ and $D$, which affects that amplitude of the underlying limit cycle. From left to right, the times are given by  $t \in \{0, 10, 50,150, 200\}$. Initial conditions are given in \cref{eqn:TravellingWaveIC} and periodic boundary conditions are used.}
\label{fig:FiveCycleTravelingWave}
\end{figure}

Repeating the same analysis for,
\begin{equation*}
    \lambda_{4,6} = \frac{5 c - \sqrt{5} \sqrt{5
   c^2-\sqrt{5} a D+a D\pm i \sqrt{2 \left(5+\sqrt{5}\right)} (a+2) D}}{10 D},
\end{equation*}
we see a similar result. Here we have,
\begin{equation*}
    c^* = \pm\frac{(a+2) \sqrt{a \left(D+\frac{3 D}{\sqrt{5}}\right)}}{2 a} \quad b^* = \mp\sqrt{\left(\sqrt{5} - 1\right) a D},
\end{equation*}
which gives a valid wave speed if and only if $ a > 0$. Again treating $\lambda_{4,6}$ as functions of wave speed, we see,
\begin{equation*}
    \lambda_{4,6}(c^*) = \pm \frac{i a}{\sqrt{5 aD \left(1+\sqrt{5}\right)} }.
\end{equation*}
Again we see a pair of pure imaginary eigenvalues emerge and clearly $\lambda'(c^*) \neq 0$ indicating a second Hopf bifurcation with emergent limit cycle implying the existence of travelling waves when $a > 0$. Thus, we have shown that the one-dimensional Vickers equation with the biased five cycle exhibits travelling waves both for $a > 0$ and $ a< 0$. We illustrate this numerically in \cref{fig:FiveCycleTravelingWave} using periodic boundary conditions and initial conditions,
\begin{equation}
    u_i(x,0) = \frac{1}{N}\left[1 + \sin\left(x - \frac{2(i-1)\pi}{N}\right) \right],
    \label{eqn:TravellingWaveIC}
\end{equation}
with $N = 5$.
That travelling waves emerge as a result of the existence of a limit cycle in \cref{eqn:TravellingODE} implies these waves are of constant amplitude, as shown in \cref{fig:FiveCycleTravelingWave}. Incidentally, this offers a numeric proof that a stable limit cycle emerges as a result of the Hopf bifurcation for certain parameter regimes in \cref{eqn:TravellingODE}. The amplitude of this limit cycle is a function of both $a$ and $D$, as shown in \cref{fig:FiveCycleTravelingWave} (middle) and (bottom). While the Hopf bifurcation is not changed by $D$, the wave speed and consequently the amplitude of the limit cycle are sensitive to this parameter. This behaviour is also noted in more complex three-strategy cases by Griffin et al. in \cite{GFW24}. From the PDE perspective, too much diffusion effectively shrinks the limit cycle to the interior fixed point, leading all solutions to approach the interior equilibrium.

We note that \cite{GFW24} originally introduced higher-order interaction terms to induce travelling waves for non-collapsing populations ($a < 0$). However, these results show that travelling waves can occur in non-collapsing populations ($a > 0$), so higher-order interactions are not necessary to ensure this occurs in the biased 5-cycle.

\subsection*{Results for the 6-cycle}
\begin{figure}[htbp]
\centering
\includegraphics[width=0.98\textwidth]{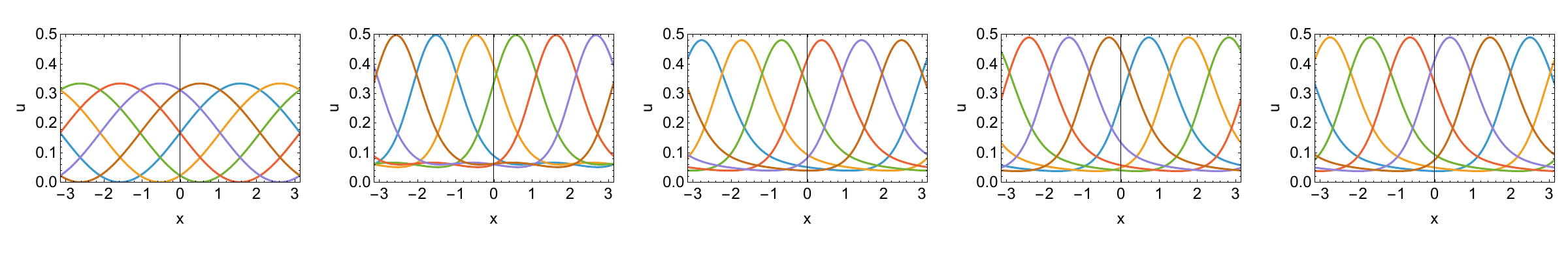}\\
\includegraphics[width=0.98\textwidth]{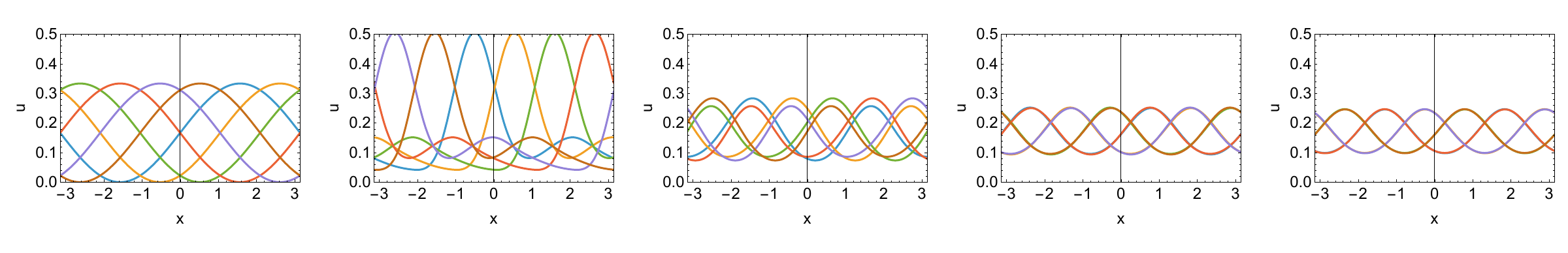}\\
\includegraphics[width=0.20\textwidth]{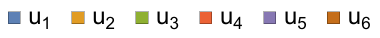}
\caption{(Top) Emergence of a travelling wave with $a = \tfrac{1}{2}$ and $D = \tfrac{1}{50}$. (Bottom) Emergence of a travelling wave with $a = -\tfrac{1}{2}$ and $D = \tfrac{1}{50}$.  From left to right, the times are given by  $t \in \{0, 10, 50,150, 200\}$. Initial conditions are given in \cref{eqn:TravellingWaveIC} and, periodic boundary conditions are used.}
\label{fig:SixCycleTravelingWave}
\end{figure}

We apply the same approach to analysing the spatial dynamics in the biased six-cycle. In this case, the eigenvalues of the Jacobian evaluated at the interior fixed point $u_i = \tfrac{1}{6}$, $v_i = 0$ for $i \in \{1,\dots,6\}$ are given by,
\begin{align}
    &\lambda_{1,2,3,4} = \frac{3c\pm\sqrt{3}\sqrt{3c^2+2aD}}{6D}\label{eqn:6evs1}\\
    &\lambda_{5,6} = \frac{3c \pm \sqrt{9c^2+3aD+3\sqrt{3}iD(2+a)}}{6D}\label{eqn:6evs2}\\
    &\lambda_{7,8} = \frac{3c \pm \sqrt{9c^2+3aD-3\sqrt{3}iD(2+a)}}{6D}\label{eqn:6evs3}\\
     &\lambda_{9,10} = \frac{3c \pm \sqrt{9c^2-3aD+3\sqrt{3}iD(2+a)}}{6D}\label{eqn:6evs4}\\
    &\lambda_{11,12} = \frac{3c \pm \sqrt{9c^2-3aD-3\sqrt{3}iD(2+a)}}{6D}\label{eqn:6evs5}
\end{align}
Using $\lambda_6$, $\lambda_8$, $\lambda_{10}$ and $\lambda_{12}$, we identify  two travelling wave speeds, 
\begin{equation}
    c^\star = \pm\frac{(2+a)k}{2\sqrt{-ak}}, 
\end{equation}
and 
\begin{equation}
    c^* = \pm\frac{(2+a)k}{2\sqrt{ak}},
\end{equation}
corresponding to positive and negative bias, as in the biased five-cycle case. As before, the eigenvalues have non-zero derivative in the bifurcation term ($c$) at the points $c = c^*$ and $c = c^\star$, thus satisfying Hopf's theorem and showing a Hopf bifurcation exists. \cref{fig:SixCycleTravelingWave} shows that the resulting limit cycles are stable, and these results prove that travelling wave solutions are possible in the 6-cycle when $a>0$ and $a<0$. We use the initial conditions given by \cref{eqn:TravellingWaveIC} with $N = 6$.


We conclude the discussion of travelling waves in the biased 6-cycle by noting an interesting phenomenon in \cref{fig:SixCycleTravelingWave}. When $a = -\tfrac{1}{2}$ the travelling wave solutions converge so that $u_{1} = u_4$, $u_{2} = u_5$, and $u_{3} = u_6$. Under this symmetry, the spatial dynamics are equivalent to the biased three-cycle studied by Griffin et al. \cite{GMD21} and the resulting travelling wave is equivalent to the one identified by them. It is unclear why these species pair together in this travelling wave solution, since there is no obvious attracting fixed point associated with this dynamic, and further study is warranted.

\subsection*{Biased 4-Cycles Do Not Admit Travelling Waves}
We contrast these results along with the known results on travelling waves in the spatial replicator generated by biased 3-cycle \cite{GMD21} the  with the biased 4-cycle. In this case, the eigenvalues of the Jacobian matrix for \cref{eqn:TravellingODE} at the fixed point $u_i = \tfrac{1}{4}$ and $v_i = 0$ for $i \in \{1,2,3,4\}$ are given by,
\begin{align}
    &\lambda_{1,2,3,4} = \frac{c \pm \sqrt{c^2+aD}}{2D}\label{eqn:4CycleEig1}\\
    &\lambda_{5,6} = \frac{c \pm \sqrt{c^2 + i(2+a)D}}{2D}\label{eqn:4CycleEig2}\\
    &\lambda_{7,8} = \frac{c \pm \sqrt{c^2 - i(2+a)D}}{2D}\label{eqn:4CycleEig3}.
\end{align}
Notice that we cannot use the approach used in the biased three \cite{GMD21}, five, and six cycle cases to show a Hopf bifurcation exists. Consequently, these dynamics do not appear to admit a closed orbit. Thus, travelling waves do not appear to exist in the dynamics generated from the biased four-cycle. 
\begin{figure}[htbp]
\centering
\includegraphics[width=0.98\textwidth]{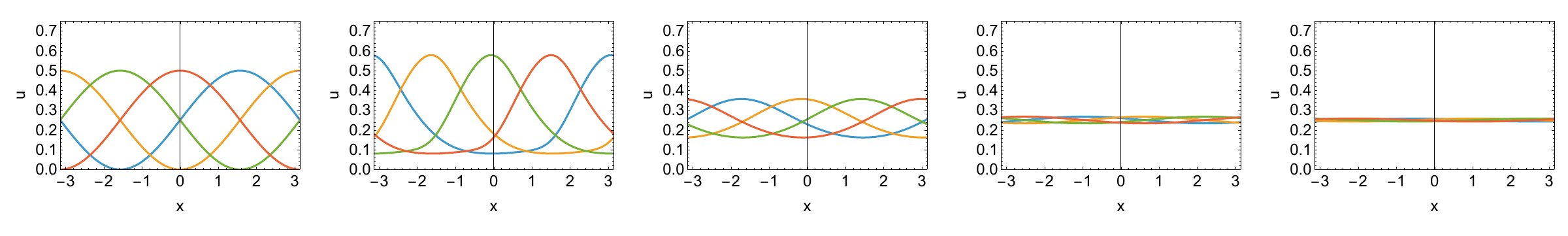}\\
\includegraphics[width=0.98\textwidth]{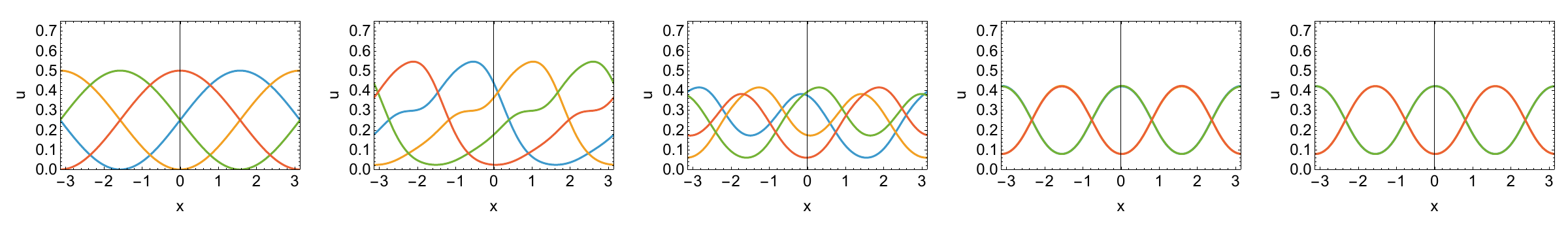}\\
\includegraphics[width=0.20\textwidth]{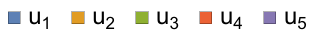}
\caption{(Top) A solution that collapses to the interior equilibrium $u_i(x,t) = \tfrac{1}{4}$ everywhere. (Bottom) Evolution to a stationary frozen (spatial) wave. Here, we see that species 1 synchronizes spatially with species 3 as does species 2 with species 4. These waves are frozen; species 1 and 3 are in phase with each other, and species 2 and 4 are in phase with each other.}
\label{fig:FourCycle}
\end{figure}
This conjecture is supported by numerical experimentation (see \cref{fig:FourCycle}), which shows that these dynamics do not exhibit constant amplitude travelling waves but do exhibit a stationary (frozen) dynamic archetypical of other biased even cycles, which we discuss in \cref{sec:FrozenWaves}.

\subsection*{Generalizations and Conjecture}
We can combine the previous analysis with the results in Griffin et al. \cite{GMD21,GFW24} to obtain a set of general conjectures about the dynamics in spatial the spatial replicator derived from the Volterra lattice.

\begin{conjecture} Consider the travelling wave ordinary differential equation, i.e., \cref{eqn:TravellingODE}, generated from a biased $n$-cycle where $n \geq 3$ and $n \neq 4$ with bias parameter $a$. If $n$ is odd and $a < 0$, then the dynamics always admit a Hopf bifurcation with a limit cycle, implying the corresponding spatial dynamics, i.e., \cref{eqn:OneDVickers}, admit a travelling wave solution. Moreover, if $ a > 0$ and $n \geq 5$, then the dynamics always admit a Hopf bifurcation with a limit cycle, implying the corresponding spatial dynamics, admit a travelling wave solution.
\end{conjecture}

Proving this conjecture is clearly left to future work, since a general proof would require showing that the resulting characteristic equations of the Jacobian matrices always admit roots with a special wave speed that leads to a Hopf bifurcation along with the construction of the first Lyapunov coefficient showing that the resulting limit cycle is stable (as in \cite{GFW24}). It is worth noting that when $a = 0$ and the aspatial dynamics are integrable, our prior analysis suggests that no travelling wave exists, which is suggestive of future work as well and would complete the analysis of travelling waves for this Volterra lattice inspired system. 

\section{Ecological Niches and Frozen Waves}\label{sec:FrozenWaves}
Consider again the travelling wave dynamics for the biased four-cycle. If $a < 0$ and $c = 0$, then from \cref{eqn:4CycleEig1}, this system admits four purely imaginary eigenvalues,
\begin{equation}
    \tilde{\lambda}_{1,2,3,4} = \frac{\pm i\sqrt{-ak}}{2D},
\end{equation}
with four additional eigenvalues with both positive and negative real components arising from the factor of the $\sqrt{\pm i}$. This suggests that the stationary dynamics may have a centre manifold and that a spatial cycle or frozen wave may be attracting for certain initial conditions.


Indeed, when this system was solved numerically with $a = -\tfrac{1}{2}$, $k = \tfrac{1}{50}$ and initial conditions given in \cref{eqn:TravellingWaveIC} are used, we discover that a frozen wave solution exists, with species one collapsing to species three and species two collapsing to species four. This is shown in \cref{fig:FourCycle} (bottom).
In ecological terms, communities form in the ecosystem in which species one and three occupy one set of communities and species two and four occupy another set of communities. This behaviour can be explained by the graph structure. Species three preys on species four, who preys on species one. Similarly, species one preys on species two, who preys on species three. Therefore, species one and three protect each other and form a niche for their mutual benefit. 

\begin{figure}[htbp]
    \centering
    \includegraphics[width=0.45\textwidth]{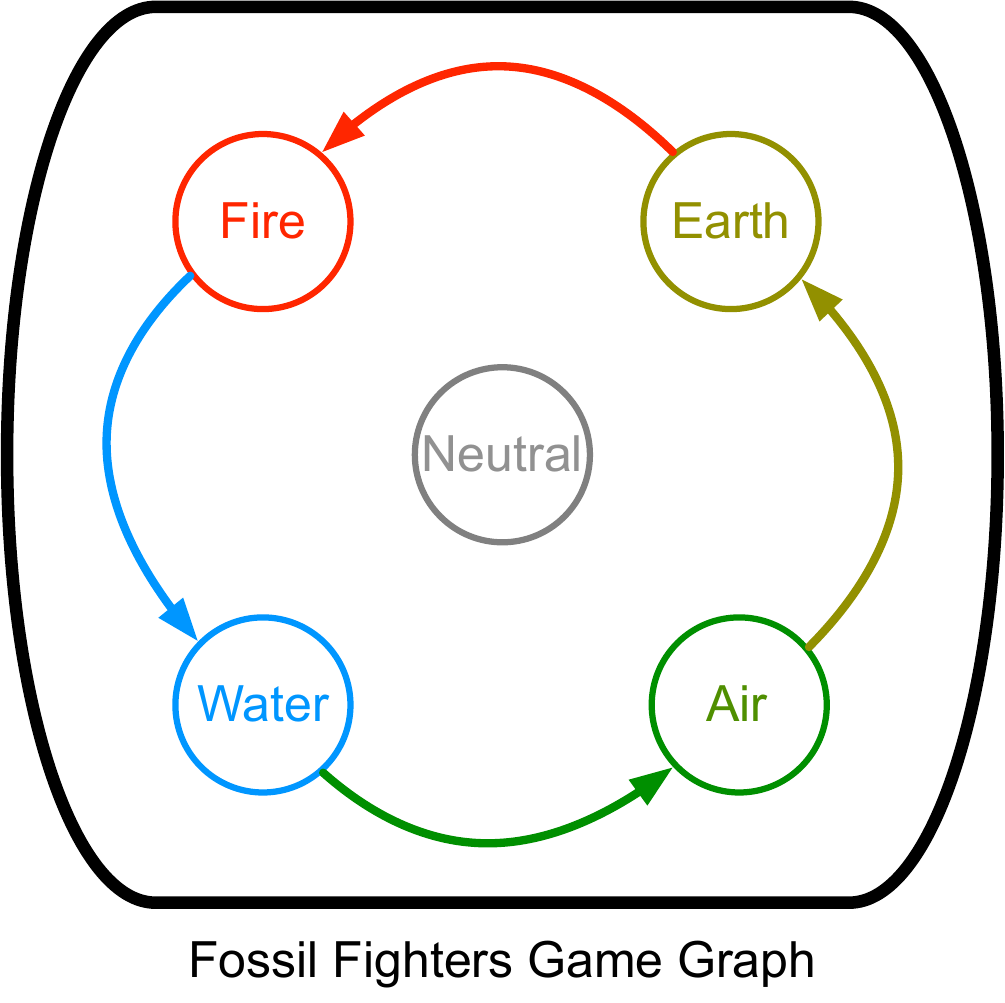}
    \caption{Graph of the type advantages within the Nintendo game Fossil Fighters. Note that this is a relabelled Four Cycle. }
    \label{fig:FossilFightersTypeAdvantages}
\end{figure}

As an aside, this dynamic appears in the popular Nintendo video game \textit{Fossil Fighters} \cite{FF25}. Gameplay consists of excavating fossils, using those fossils to revive dinosaurs, and using those dinosaurs to fight battles in teams of three. Each dinosaur can have a `type', which influences how effective it is against opponents. Types and advantages are shown in \cref{fig:FossilFightersTypeAdvantages}. 
One successful strategy when constructing teams consists of pairing two Earth-type dinosaurs with a Water-type dinosaur so that the team as a whole can effectively defend itself against both Fire-type and Air-type dinosaurs, thus mimicking ecological communities. 


\subsection{Turing Pattern Analysis}
\begin{figure}[htbp]
    \centering
    \includegraphics[width=0.45\linewidth]{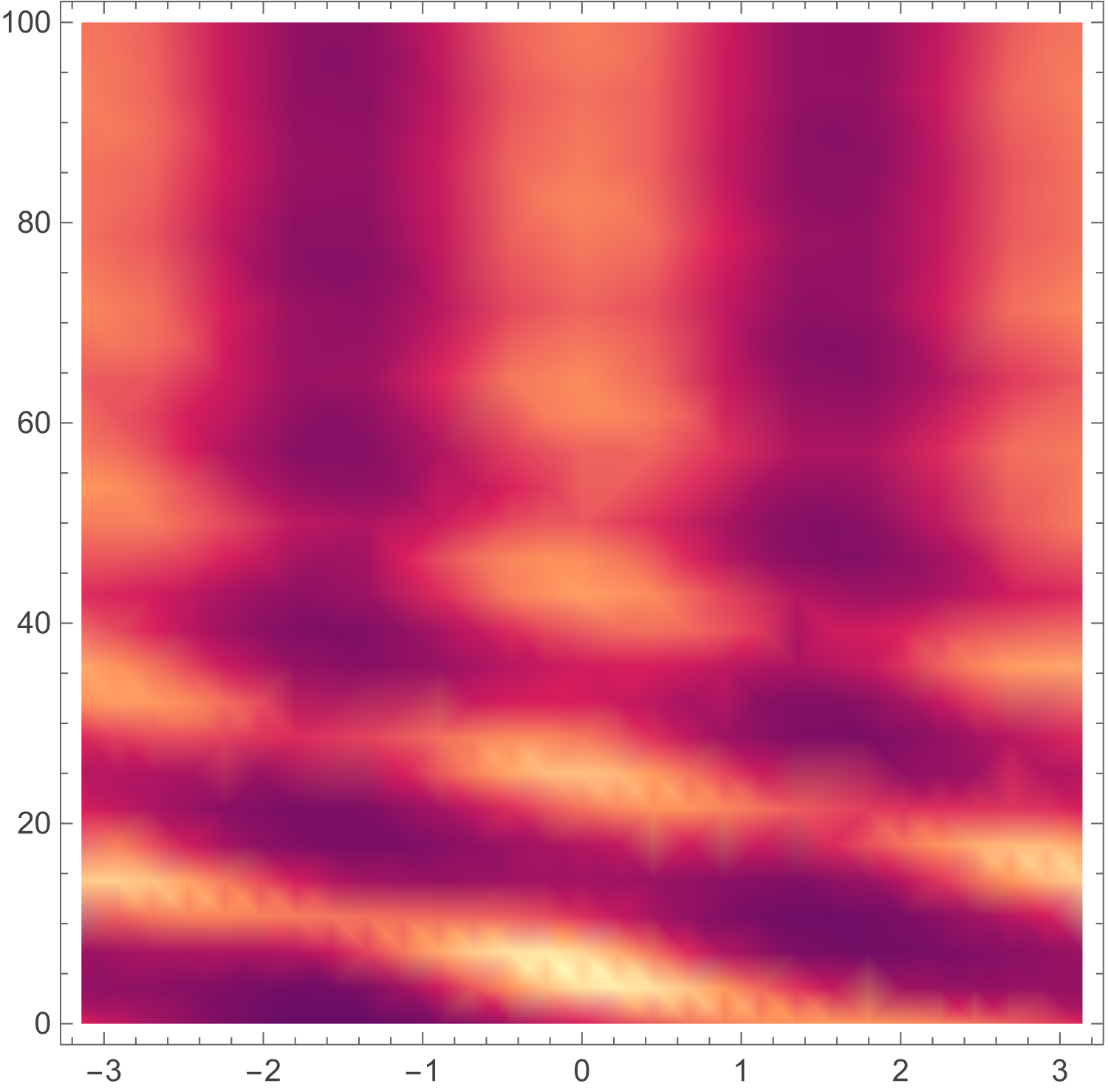}\quad
    \includegraphics[width=0.45\linewidth]{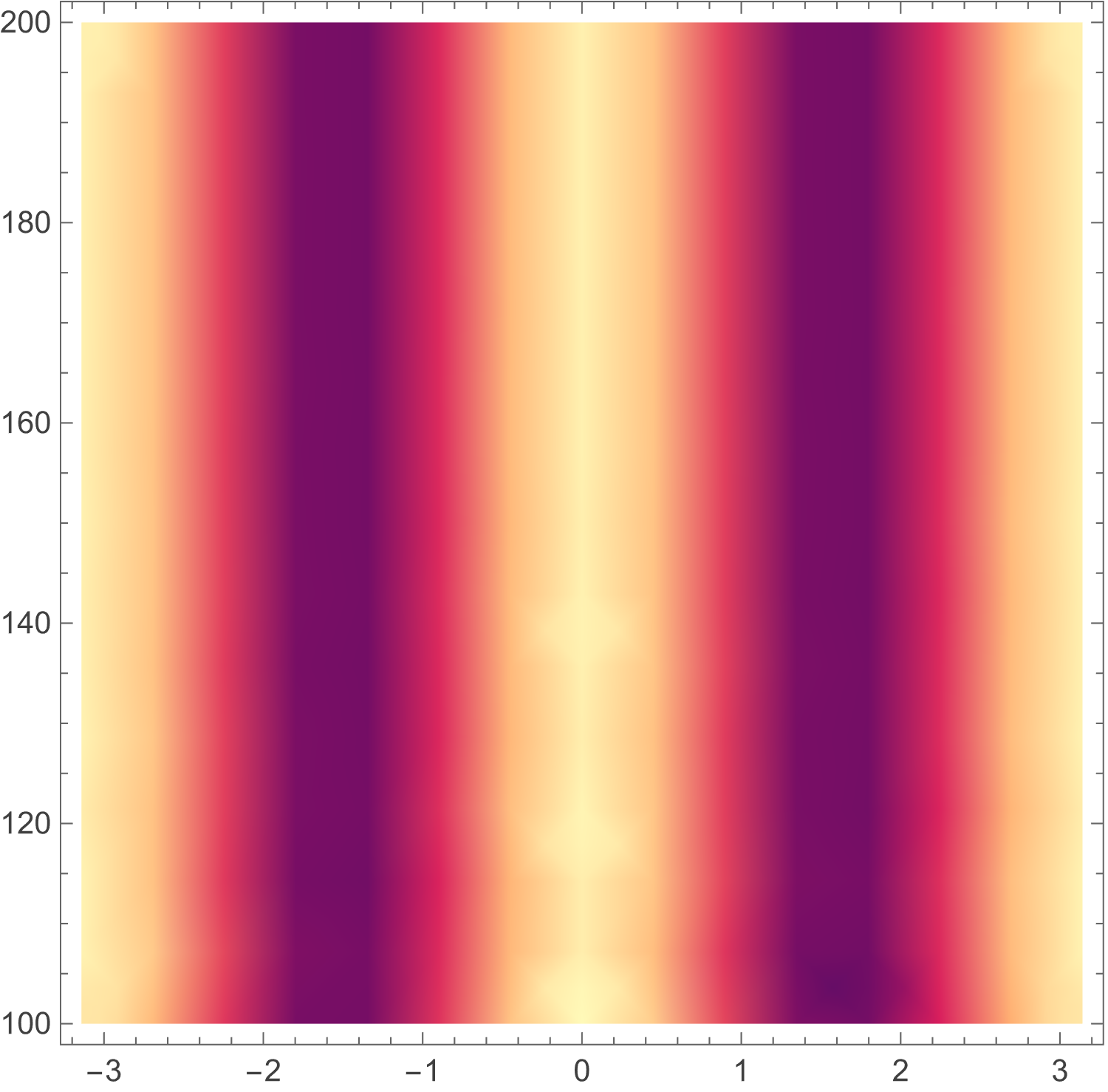}
    \caption{This is a density plot of species 1 in space over time. This striped pattern looks like a Turing Pattern. }
    \label{fig:4CycleDensityPlot}
\end{figure}

Consider the density plot in \cref{fig:4CycleDensityPlot} for the biased four-cycle with $a < 0$. 
In this case, we know that in the aspatial dynamics, the interior fixed point is unstable, but there are (apparently) semi-stable boundary points that could explain the observed Turing-like patterns. To determine if these are Turing patterns  resulting from diffusion-driven instabilities within the model, we follow Murray's algorithm \cite[Chapter 2]{murray2007mathematical}. Specifically, we find the roots of the characteristic polynomial corresponding, 
\begin{equation}
    |(\lambda + Dk^2)\textbf{I} - \textbf{J}|=0
    \label{eqn:Turing Pattern algorithm}
\end{equation}
where $D$ is the diffusion constant, $k$ is the wave number, and $\textbf{J}$ is the Jacobian of the reaction dynamics in \cref{eqn:OneDVickers} evaluated at an appropriate fixed point. At the interior fixed point, the values of $\lambda$ satisfying \cref{eqn:Turing Pattern algorithm} are,
\begin{align*}
    &\lambda_{1,2} = -\frac{a}{4} - Dk^2\\
    &\lambda_{3,4} = -Dk^2 \pm i\left(\frac{1}{2} + \frac{a}{4}\right).
\end{align*}
Thus, the diffusion term is always stabilizing this system. This holds true at the other effectively stable fixed points on the boundary, see \cref{fig:FourCycleExample} (right), showing that these patterns are not Turing patterns in the classic sense; i.e., diffusion is not destabilizing a stable fixed point.




\subsection{Frozen Waves in Higher Order Even cycles}

After observing frozen waves as stationary solutions to the spatial, biased 4-cycle dynamics, we postulated that stationary solutions of this type could occur in the spatial dynamics of all biased even cycles when the bias was negative (i.e., $a < 0$). Specifically, we predicted that species would form two different niches composed of the odd-numbered species and the even-numbered species. Forming these niches would ensure mutual protection for every species within the niche, since odd-numbered species consume and are consumed by even-numbered species and vice versa. 

\begin{figure}[htbp]
\centering
\includegraphics[width=0.45\textwidth]{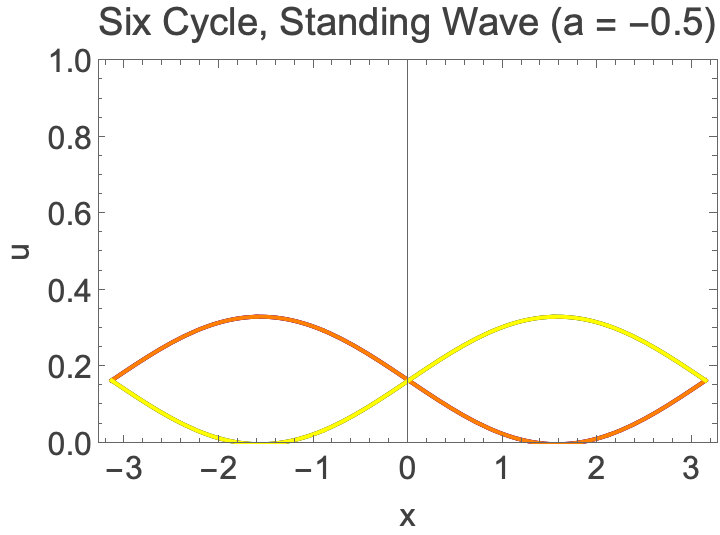}\quad
\includegraphics[width=0.45\textwidth]{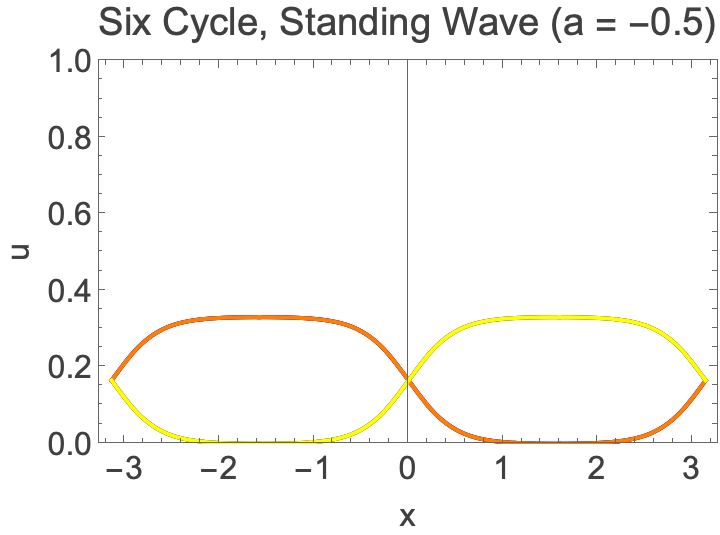}
\caption{(Left) Initial condition used for the biased 6-cycle to induce a frozen wave solution. (Right) Final frozen wave solution. Note the slight differences between the initial condition and the final solution.}
\label{fig:Sixcyclestandingwave}
\end{figure}

\begin{figure}[htbp]
\centering
\includegraphics[width=0.45\textwidth]{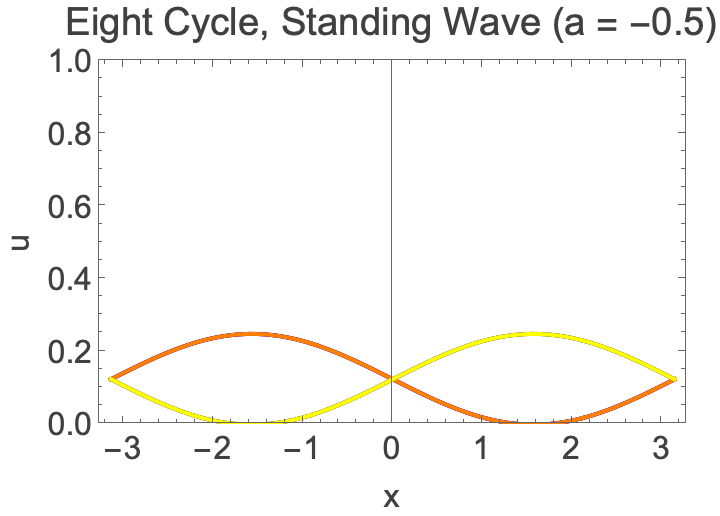}\quad
\includegraphics[width=0.45\textwidth]{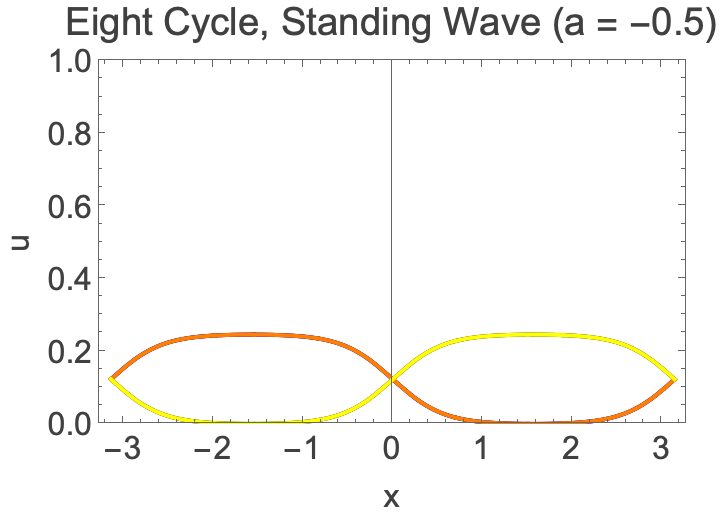}
\caption{(Left) Initial condition used for the 8-cycle to induce a frozen wave solution. (Right) Final frozen wave solution. Note the slight differences between the initial condition and the final solution.}
\label{fig:Eightcyclestandingwave}
\end{figure}

As evidence for this, consider the first four eigenvalues of the biased 6-cycle in \cref{eqn:6evs1}. When $c = 0$ and $a < 0$ we have,
\begin{equation*}
\tilde{\lambda}_{1,2,3,4} = \pm i\frac{\sqrt{-6aD}}{6D},
\label{eqn:6CentreManifold}
\end{equation*}
suggesting a centre manifold in the degenerate ($c = 0$) travelling wave dynamics for $a < 0$, just as in the biased 4-cycle case. As we have already demonstrated in \cref{fig:SixCycleTravelingWave}, initial conditions of the form given in \cref{eqn:TravellingWaveIC} lead to travelling waves. However, if we seed ecological communities with the initial conditions,
\begin{equation}
    u_{i}(x,0) = 
  \begin{cases}
       \frac{1}{6}(1+\sin(x)) ,& \text{if $i$ is odd}\\
      \frac{1}{6} (1-\sin(x)), & \text{if $i$ is even},
    \end{cases}
    \label{initial conditions}
\end{equation}
then the system maintains these ecological communities (with minor changes) over time. This is shown in \cref{fig:Sixcyclestandingwave} with $a = -\tfrac{1}{2}$ and $D = \tfrac{1}{100}$.
Therefore, frozen wave solutions (ecological niches) can form a stationary solution in the biased 6-cycle. 
Applying a similar approach with the biased $8$-cycle using the initial conditions,
\begin{equation}
    u_{i} = 
    \begin{cases}
        \frac{1}{8}(1+\sin(x)) ,& \text{if $i$ is odd}\\
       \frac{1}{8} (1-\sin(x)), & \text{if $i$ is even},
    \end{cases}
    \label{initial conditions2}
\end{equation}
also generates frozen waves with $a = -\tfrac{1}{2}$ and $D = \tfrac{1}{100}$, as shown in \cref{fig:Eightcyclestandingwave}.

As in the case of the 4-cycle, these stationary behaviours are inconsistent with Turing patterns. Thus, the results shown for the biased 4-, 6-, and 8- cycle lead us to the following conjecture.
\begin{conjecture}
Consider the one-dimensional spatial replicator, i.e., \cref{eqn:OneDVickers} with reaction term generated by a $2n$-cycle with $n \geq 2$. Assuming periodic boundary conditions and initial conditions of the form,
\begin{equation}
    u_{i}(x,0) = 
    \begin{cases}
        \frac{1}{2n}(1+\sin(x)) ,& \text{if $i$ is odd}\\
       \frac{1}{2n} (1-\sin(x)), & \text{if $i$ is even},
    \end{cases}
    \label{initial conditions3}
\end{equation}
the dynamics will evolve to a frozen wave stationary solution. That is, ecological niches will be maintained. 
\end{conjecture}


\subsection{Random Start in the Biased 4- and 6- Cycles}

\begin{figure}[htbp]
\centering
\includegraphics[width=0.31\textwidth]{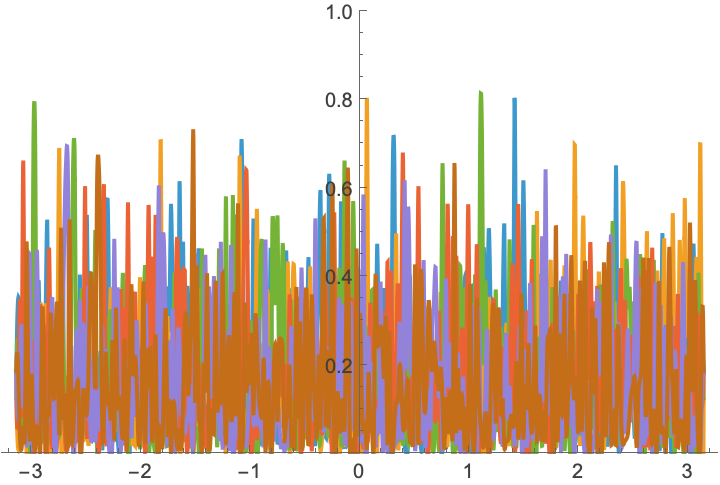}\quad
\includegraphics[width=0.31\textwidth]{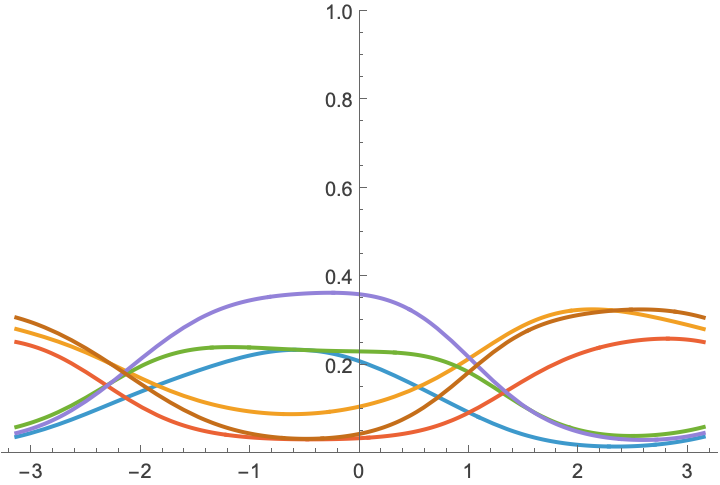}\quad
\includegraphics[width=0.31\textwidth]{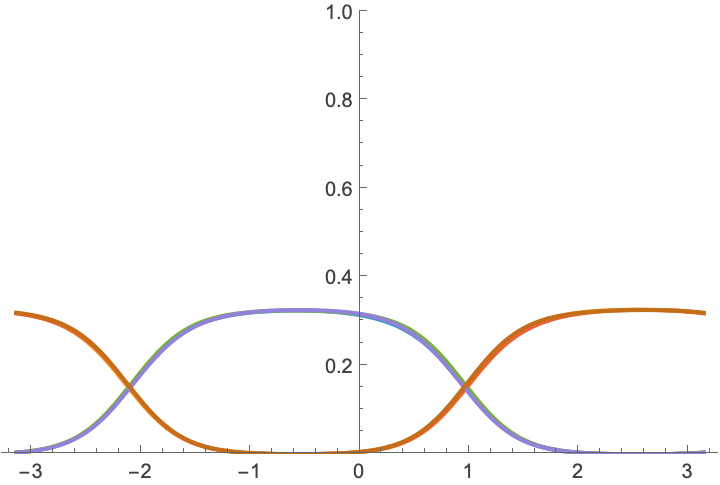}
\caption{(Left) A random initial condition for the biased 6-cycle dynamics. (Middle) Intermediate time step where organization arises from randomness. Note that the ecological communities are starting to form. (Right) Final state of the system, where the ecological communities are fully formed.\ref{fig:Sixcyclestandingwave}.}
\label{fig:1DRandomIC}
\end{figure}

While we determined that frozen waves were possible stationary solutions of the spatial dynamics of the biased 4-, 6-, and 8- cycles, we cannot determine whether these states are attracting in any sense (except in the $4$-cycle) because our initial conditions already lead to solutions exhibiting this behaviour. To explore whether this class of stationary solution has a non-trivial basin of attraction, we use a random initial condition and numerically integrate over a sufficient time horizon. We used the spatial domain $[-\pi, \pi]$ with $0 \leq t \leq 400$ and used periodic boundary conditions (solutions on the circle). To generate the initial condition, we broke the spatial interval into 400 discrete points and chose a random point in $\Delta_n$ to be assigned to each point. We forced the periodic boundary condition to hold by setting the end points to be equal to their average. We then interpolated these points into $n-1$ functions $f_1(x),\dots,f_{n-1}(x)$ and set the $n^\text{th}$ initial condition to be $f_n(x) = 1 - f_1(x) - \cdots - f_{n-1}(x)$. Mathematica code for this is available in the SI. Example initial conditions, evolution and an example final frozen wave (ecological communities) are shown in \cref{fig:1DRandomIC}.
%
%


We used 64 random initial conditions for both a $4$-cycle and $6$-cycle and manually determined whether a frozen wave resulted. We varied the bias using $a=-\tfrac{1}{2}$ and $a = -\tfrac{1}{3}$. We varied the diffusion constant using $D = \tfrac{1}{100}$ and $D = \tfrac{1}{75}$. Output plots are shown in \cref{fig:SixtyFour4,fig:SixtyFour6,fig:SixtyFour4HiD,fig:SixtyFour6HiD,fig:SixtyFour4LoA,fig:SixtyFour6LoA}. Results are summarized in \cref{tab:Results} where $95\%$ confidence intervals on the proportion of random initial conditions that resulted in frozen waves (ecological communities) are provided. 
\begin{table}[htbp]
\centering
\begin{tabular}{|l|c|c|c|c|}
\hline
\textbf{Graph} & $D$ & $a$ & \textbf{Frozen Wave Count} & \textbf{Prop. $\pm$ CI}\\
\hline
4-Cycle & $0.01$ & -0.5 & $34$ & $0.53\pm0.12$\\
\hline
4-Cycle & $0.0133$ & -0.5 &  $30$ & $0.48\pm0.12$\\
\hline
4-Cycle & $0.01$ & -0.33 & $25$ & $0.39\pm0.12$\\
\hline
6-Cycle & $0.01$ & -0.5& $23$ & $0.35\pm0.12$\\
\hline
6-Cycle & $0.0133$ & -0.5 & $28$ & $0.43\pm0.12$\\
\hline
6-Cycle & $0.01$ & -0.33 & $20$ & $0.31\pm0.11$\\
\hline
\end{tabular}
\caption{Table showing the proportion of integrations from a random initial condition that resulted in a frozen wave (ecological niches).}
\label{tab:Results}
\end{table}
We observe that it is possible frozen wave (ecological niche) formation is sensitive to the diffusion and bias constants, though there is no clear pattern in this small experiment. However, it is worth noting that frozen waves (ecological communities) formed in all cases. We also note that when communities did not form, non-travelling wave oscillating solutions seemed to result.  

\begin{figure}[htbp]
\centering
\includegraphics[width=0.85\textwidth]{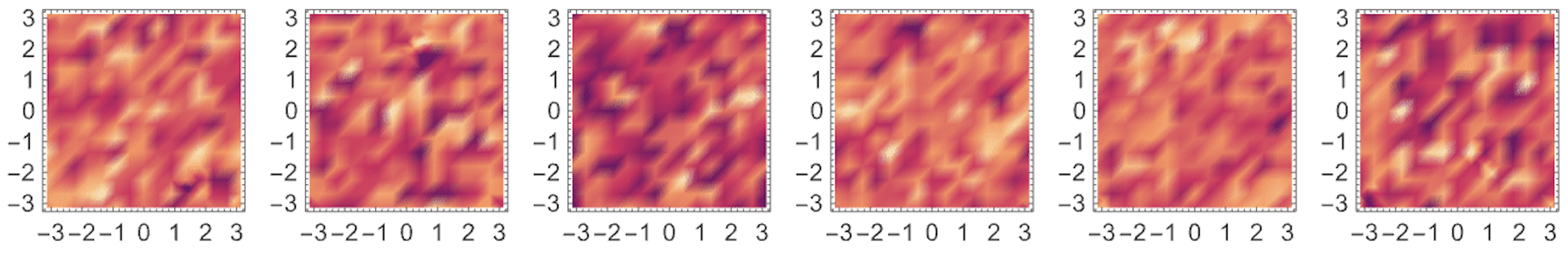}\\
\includegraphics[width=0.85\textwidth]{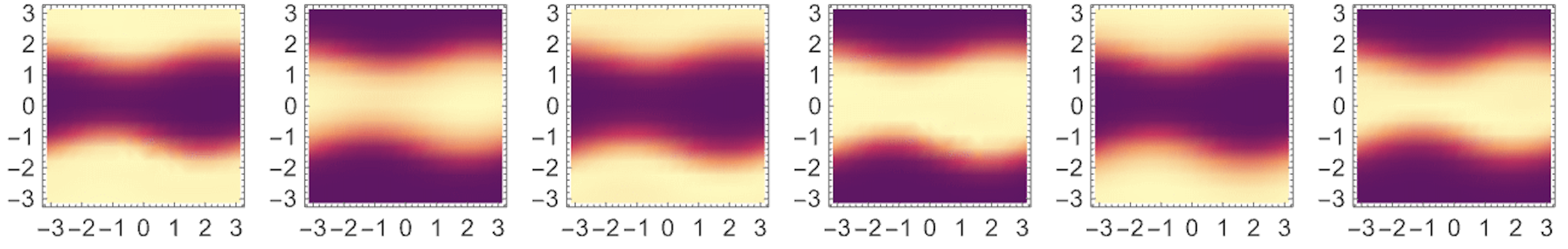}\\
\includegraphics[width=0.85\textwidth]{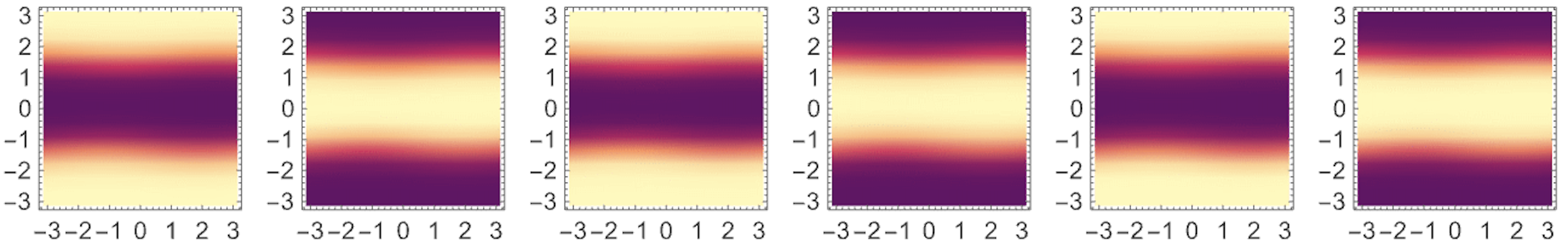}
\caption{(Top) A random initial condition for the six cycle. Species $1$ - $6$ are from left to right. (Middle) Intermediate time step where organization arises from randomness. That is, the ecological communities are starting to form. (Right) Final state of the system, where the ecological communities have fully formed.}
\label{fig:2DRandomIC}
\end{figure}

Finally, we performed an integration from a random two-dimensional initial conditions as well using periodic boundary conditions in the biased 6-cycle. Like Turing patterns, which are known to be independent of dimension, it appears these communities will form in higher dimensions as illustrated in \cref{fig:2DRandomIC}.
Further exploration of this phenomenon is left for future work but prompts the obvious conjecture that all biased $2n$-cycles admit these frozen wave solutions or ecological niche solutions from a positive measure set of initial conditions. A formal definition of these solutions (which are not always conveniently periodic, but have wave-like properties) and a precise specification of this conjecture is clearly required and will be a part of future work.


\section{Conclusion}\label{sec:Conclusion}

In this paper, we defined the biased Volterra lattice (as an extension of biased rock-paper-scissors explored extensively in evolutionary game theory. We extended work in \cite{GMD21,GFW24}, which studied with the three-cycle game (rock-paper-scissors) to show that travelling wave solutions are possible in the spatial replicator equations in higher order cycles. We proved that travelling wave solutions exist in the 5- and 6- cycles, and they were possible with both positive and negative bias terms. These results show agreement between the replicator equations and ecological observations of natural travelling waves in populations, suggesting their potential viability in future ecological modelling. Additionally, we identified a new class of stationary solutions within the spatial replicator: frozen waves. These stationary solutions appear only in dynamics generated by even cycles using negative bias terms and consist of two ecological niches, one composed of even species and the other composed of odd species. These niches are periodic in space but constant in time, and appear to exist one-half period out of phase with each other. We used numerical methods to show that this class of stationary solution seems to occur with statistically significant frequency when these systems are initialized randomly. Because Turing patterns were ruled out, these seem to be novel examples of a new type of pattern formation. 

There are several future directions for this work. Two conjectures were proposed in this paper, and proving them is of obvious interest. Furthermore, previous work by Visomirski and Griffin \cite{VG24} showed that there is a link between graph structure and the aspatial replicator dynamics. It is obviously worth asking whether a similar relationship could be determined in both the spatial case and the aspatial case with the addition of the bias terms used in this paper.



\bibliographystyle{iopart-num}
\bibliography{main}

\appendix

\section{Numerical Results in Community Formation}
Numerical results on 128 numerical experiments starting both the spatial biased 4-cycle and biased 6-cycle from random conditions are shown in the figures below.

\begin{figure}[htbp]
\includegraphics[width=0.95\textwidth]{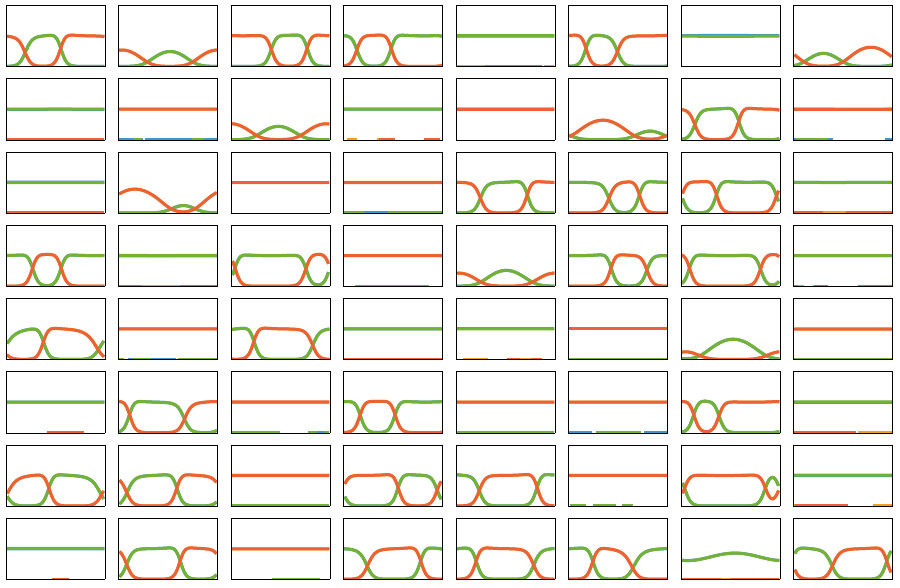}
\caption{Sixty-four plots of four-cycle evolution from a random initial condition. Thirty-four show community formation. The others exhibit oscillating behaviour. Here, $a = -\tfrac{1}{2}$ and $D = \tfrac{1}{100}$}
\label{fig:SixtyFour4}
\end{figure}

\begin{figure}[htbp]
\includegraphics[width=0.95\textwidth]{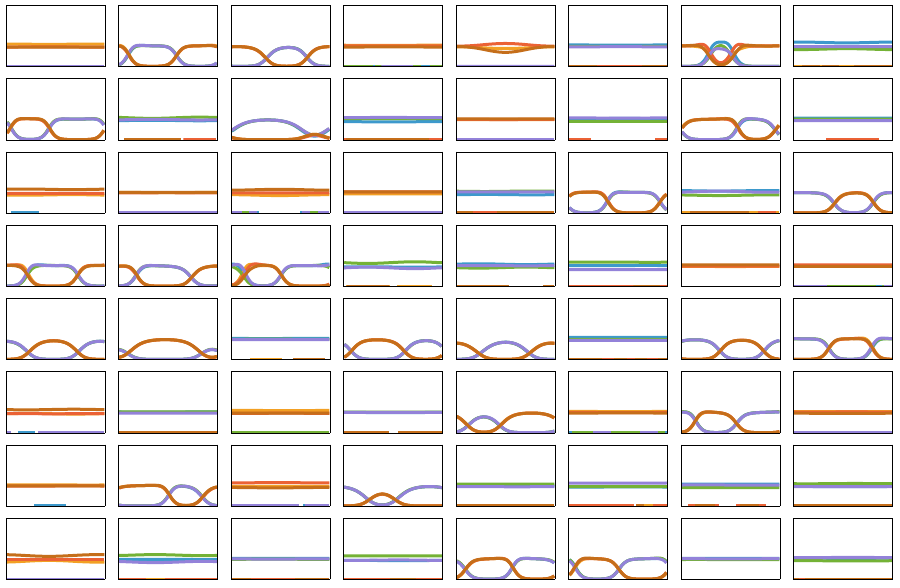}
\caption{Sixty-four plots of six-cycle evolution from a random initial condition. Twenty-three show community formation. The others exhibit oscillating behaviour. Here, $a = -\tfrac{1}{2}$ and $D = \tfrac{1}{100}$}
\label{fig:SixtyFour6}
\end{figure}

\begin{figure}[htbp]
\includegraphics[width=0.95\textwidth]{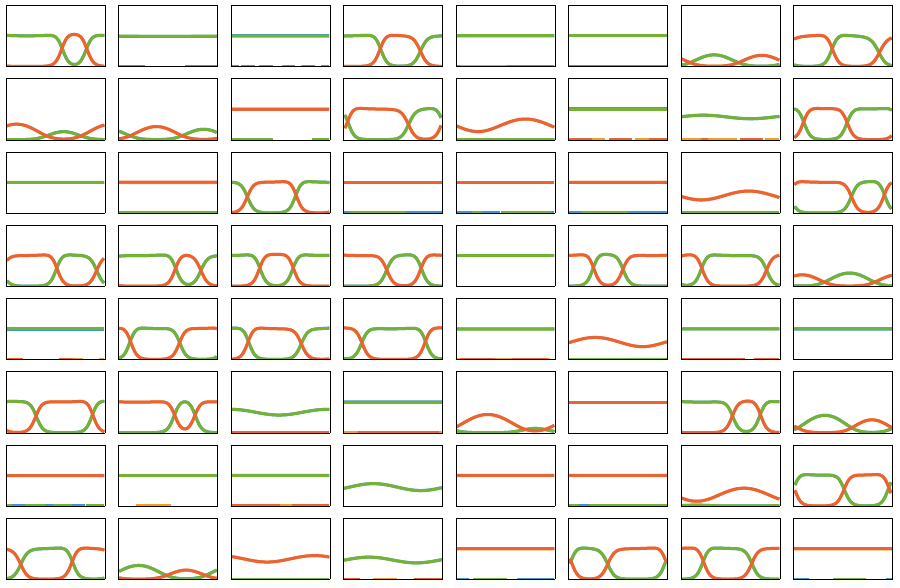}
\caption{Sixty-four plots of four-cycle evolution from a random initial condition. Thirty show community formation. The others exhibit oscillating behaviour. Here, $a = -\tfrac{1}{2}$ and $D = \tfrac{1}{75}$}
\label{fig:SixtyFour4HiD}
\end{figure}

\begin{figure}[htbp]
\includegraphics[width=0.95\textwidth]{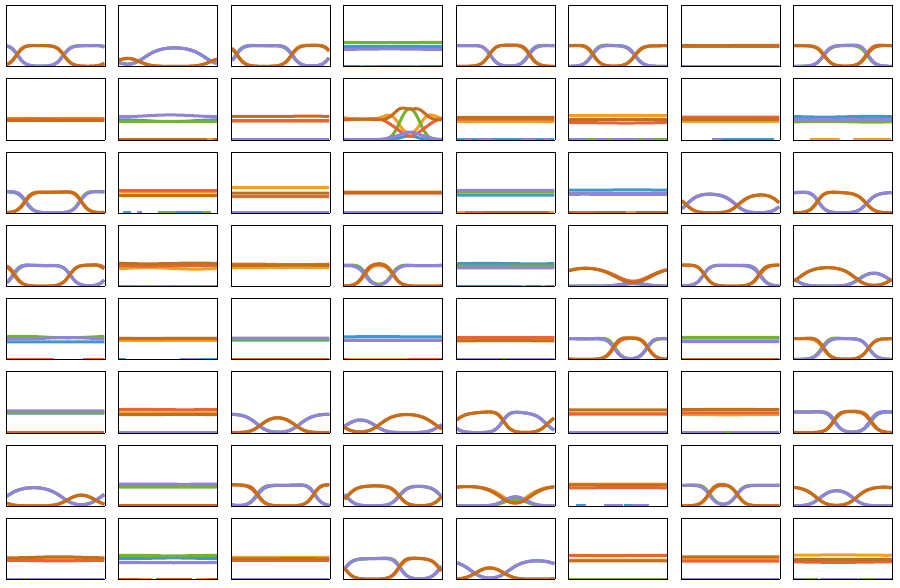}
\caption{Sixty-four plots of six-cycle evolution from a random initial condition. Twenty-eight show community formation. The others exhibit oscillating behaviour. Here, $a = -\tfrac{1}{2}$ and $D = \tfrac{1}{75}$}
\label{fig:SixtyFour6HiD}
\end{figure}

\begin{figure}[htbp]
\includegraphics[width=0.95\textwidth]{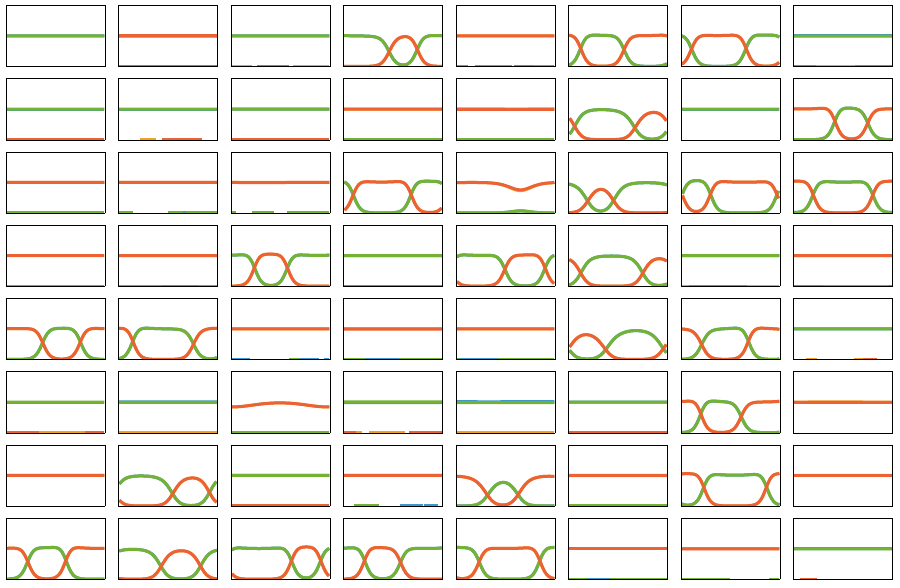}
\caption{Sixty-four plots of four-cycle evolution from a random initial condition. Twenty-five show community formation. The others exhibit oscillating behaviour. Here, $a = -\tfrac{1}{2}$ and $D = \tfrac{1}{75}$}
\label{fig:SixtyFour4LoA}
\end{figure}

\begin{figure}[htbp]
\includegraphics[width=0.95\textwidth]{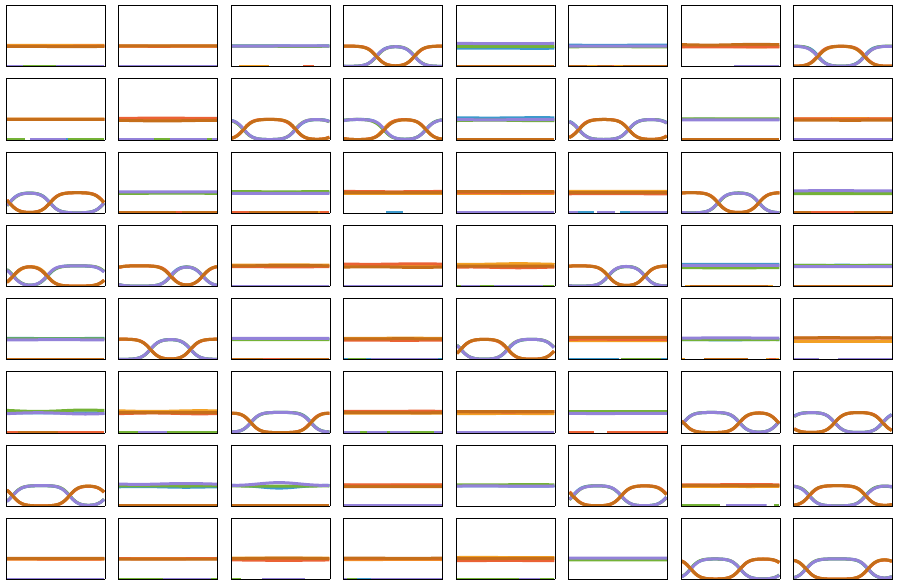}
\caption{Sixty-four plots of six-cycle evolution from a random initial condition. Twenty show community formation. The others exhibit oscillating behaviour. Here, $a = -\tfrac{1}{2}$ and $D = \tfrac{1}{75}$}
\label{fig:SixtyFour6LoA}
\end{figure}

\end{document}